\documentclass[reprint,superscriptaddress,amsmath,amssymb,aps,longbibliography]{revtex4-2}

\usepackage{amsmath}
\usepackage{amssymb}
\usepackage{amsthm}
\usepackage{amsfonts}
\usepackage[caption=false]{subfig}
\usepackage{hyperref}
\usepackage{tikz}
\usepackage{relsize}
\usepackage{color,soul}
\usepackage[utf8]{inputenc}
\usepackage{capt-of}
\usepackage{mathtools}
\usepackage{float}
\usepackage[section]{placeins}
\usepackage{listings}
\usepackage{svg}
\usepackage{siunitx}
\usepackage{hhline}
\usepackage[english]{babel}

\usepackage[T1]{fontenc}  
\usetikzlibrary{decorations.pathreplacing}
\usepackage{braket}

\theoremstyle{definition}

\theoremstyle{definition}

\theoremstyle{definition}

\newcommand{\eq}[1]{\hyperref[eq:#1]{Equation~\ref*{eq:#1}}}
\renewcommand{\sec}[1]{\hyperref[sec:#1]{Section~\ref*{sec:#1}}}
\newcommand{\fig}[2][]{\hyperref[fig:#2]{Fig.~\ref*{fig:#2}\textbf{#1}}}
\newcommand{\tbl}[1]{\hyperref[tbl:#1]{Table~\ref*{tbl:#1}}}
\newcommand{\theoremref}[1]{\hyperref[theorem:#1]{Theorem~\ref*{theorem:#1}}}
\newcommand{\definitionref}[1]{\hyperref[definition:#1]{Definition~\ref*{definition:#1}}}

\newcommand{\suppsec}[1]{(see SM sec. #1)}

\begin{document}

\title{Demonstrating dynamic surface codes}

\date{\today}
\author{Google Quantum AI and Collaborators}

\begin{abstract}

A remarkable characteristic of quantum computing is the potential for reliable computation despite faulty qubits. This can be achieved through quantum error correction, which is typically implemented by repeatedly applying static syndrome checks, permitting correction of logical information~\cite{kitaev_fault-tolerant_1997, fowler_surface_2012, terhal_quantum_2015}.
Recently, the development of time-dynamic approaches to error correction has uncovered new codes~\cite{hastings_dynamically_2021} and new code implementations~\cite{mcewen_relaxing_2023, gidney_new_2023, shaw_lowering_2024}. 
In this work, we experimentally demonstrate three time-dynamic implementations of the surface code, each offering a unique solution to hardware design challenges and introducing flexibility in surface code realization.
First, we embed the surface code on a hexagonal lattice, reducing the necessary couplings per qubit from four to three.
Second, we \emph{walk} a surface code, swapping the role of data and measure qubits each round, achieving error correction with built-in removal of accumulated non-computational errors. 
Finally, we realize the surface code using iSWAP gates instead of the traditional CNOT, extending the set of viable gates for error correction without additional overhead.
We measure the error suppression factor when scaling from distance-3 to distance-5 codes of $\Lambda_{35, \text{hex}} = 2.15(2) $, $\Lambda_{35, \text{walk}} = 1.69(6)$, and $\Lambda_{35, \text{iSWAP}} = 1.56(2)$, achieving state-of-the-art error suppression for each. 
With detailed error budgeting, we explore their performance trade-offs and implications for hardware design.
This work demonstrates that dynamic circuit approaches satisfy the demands for fault-tolerance and opens new alternative avenues for scalable hardware design~\cite{debroy_luci_2024}.

\end{abstract}

\maketitle

\setlength{\textfloatsep}{10pt}
\setlength{\floatsep}{2pt}
\setlength{\intextsep}{2pt}

Quantum Error Correction (QEC) enables accurate quantum computation in the presence of uncontrolled physical noise \cite{shor_scheme_1995}.
The surface code~\cite{kitaev_fault-tolerant_1997, bravyi_quantum_1998, dennis_topological_2002} is a preeminent example, capable of suppressing physical errors below a relatively high threshold and providing excellent capabilities for logical computation~\cite{fowler_surface_2012, horsman_surface_2012}. 
These desirable characteristics have inspired extensive experimental work on the surface code~\cite{krinner_realizing_2022, google_quantum_ai_suppressing_2023}, including recently demonstrated below-threshold performance on a superconducting quantum processor~\cite{google_quantum_ai_quantum_2025} and transversal logical gates in a neutral atom quantum processor~\cite{bluvstein_quantum_2022}. 
To date, all of these experiments have been designed around a square grid of statically assigned qubits implementing CNOT/CZ entangling interactions to match the standard circuit implementation.
This was assumed to comprise the least demanding experimental implementation of the code, since fault-tolerant circuits are traditionally constructed from static QEC codes.
\begin{figure}[bt!]
    \centering
    \includegraphics{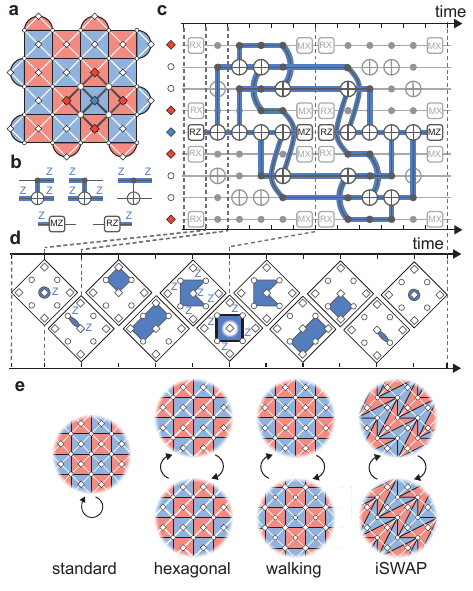}
    \setlength{\abovecaptionskip}{-5pt}  
    \setlength{\belowcaptionskip}{-5pt} 
    \caption{\textbf{Dynamic implementation of quantum error correction.} 
    \textbf{a}: A distance-5 surface code on a square lattice. The colors indicates the stabilizer Pauli flavor (Red=X, Blue=Z).
    \textbf{b}: Propagation rules for the Z-type detecting region.
    \textbf{c}: A Z-type detection region evolving through the circuit for the standard implementation, highlighting where Pauli errors that anti-commute will be detected. 
    \textbf{d}: Time slices of the detecting region between gates layers of (\textbf{c}).
    \textbf{e}: End-cycle (after measurement) time-slices for four surface code implementations: standard, along with hexagonal, walking, and iSWAP, which feature alternating end cycle states. 
    }
    \label{fig:intro}
\end{figure}

Recent progress in the theory of time-dynamic QEC, including dynamic codes~\cite{hastings_dynamically_2021} and space-time stabilizers~\cite{gottesman_opportunities_2022, delfosse_spacetime_2023}, challenge the traditional static stabilizer approach and unlock new opportunities for the construction of fault-tolerant circuits. 
The new framework of the space-time \emph{detecting region} generalizes these approaches, defined by local and connected regions of space-time that report the presence of errors.
This framework permits the re-interpretation of QEC circuits as intermeshed detecting regions, featuring overlapping sensitivity such that all physical errors can be appropriately triangulated. 
It also highlights that maintaining the precise shape of such detecting regions is less crucial than ensuring all circuit locations are covered. 
Exploiting this freedom opens the door to an expansive equivalence class of circuits for implementing well-understood QEC codes~\cite{mcewen_relaxing_2023, bombin_unifying_2023, gidney_new_2023, camps_leakage_2024, shaw_lowering_2024}, providing distinct flexibility in their hardware implementation.  

In this work, we demonstrate the ability to switch out the circuit implementation of the surface code, achieving error suppression from distance-3 to distance-5 with three time-dynamic circuits that feature qualitatively different demands on the underlying hardware.
These circuits are examples representing particularly large departures from previous assumptions about the hardware and gateset necessary to implement the surface code. 
First, we embed the surface code onto a \emph{hexagonal} grid, reducing the necessary qubit connectivity from four neighbours to three.
Second, we perform a \emph{walking} circuit in which the qubits are dynamically reassigned between data and measure, frustrating the spread of time-correlated errors.
Finally, we perform the surface code using \emph{iSWAP} gates rather than CNOT/CZ gates, demonstrating that QEC can be performed using entangling interactions previously reserved for NISQ experiments.
We operate all three implementations on Willow processors~\cite{google_quantum_ai_quantum_2025}, optimised for the standard surface code circuit, permitting direct comparisons of performance but leaving open the possibility of optimising future hardware around alternative circuits.

The newfound freedom in circuit implementation can be understood naturally using detecting regions, which indicate all locations in a QEC circuit where a non-commuting error would be detected by measurements the region touches.
\fig[a]{intro} shows a standard distance-5 surface code patch, where we highlight a single detecting region in \fig[c]{intro}.
This Z-type detecting region is sensitive to any X or Y error occurring in highlighted locations in the circuit.
The region propagates through the circuit in the same way that Pauli terms propagate through Clifford operations, illustrated in \fig[b]{intro}, with more details given in \cite{mcewen_relaxing_2023}.
The detecting region persists over two consecutive cycles, first expanding from a single qubit to cover the static code stabilizer immediately after the first measurement gate, which we call the \emph{end-cycle state}, then contracting to end in a second measurement.
Time-slices of the detecting region  are shown in \fig[d]{intro};
using a tiling of spacetime through overlapping detecting regions, we ensure that all relevant circuit locations are checked for errors.
Given this general framework, circuit flexibility arises by modifying the gate layers and associated detecting regions while preserving the tiling of spacetime.

The end-cycle states for the three alternative constructions measured in this work are illustrated in \fig[e]{intro}. 
Each alternative construction features unique alternating end-cycle states, as opposed to the standard code which has a static end-cycle state. 
Such periodic deformation of the detecting region tiling allows the alternative circuit constructions.
In this manuscript, we measure the QEC performance of these three time-dynamic circuits in isolation using a logical memory experiment, proving their ability to reduce the logical error rate as the surface code distance is increased from 3 to 5, while discussing their unique hardware requirements. 
\vspace{-3mm}
\section*{Hexagonal lattice Surface Code}
\vspace{-1mm}
\begin{figure}
    \centering
    \includegraphics{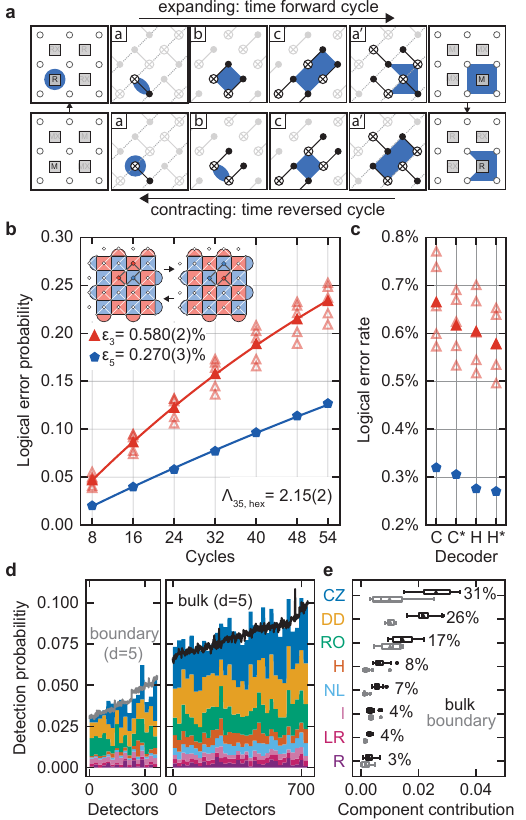}
    \caption{\textbf{Hexagonal lattice surface code.} 
    \textbf{a}: Z-type detection region time-slices. 
    Each panel shows a gate layer and detecting region after. 
    \textbf{b}: Measured logical error probabilities for a distance-5 and five constituent distance-3 hexagonal lattice surface codes with decoder H*.
    \textbf{c}: Comparison of decoders: correlated matching (C) and Harmony (H) (star indicates a reinforcement learning prior).
    \textbf{d}: Detection probability budget.
    Solid grey and black lines are measured boundary and bulk detection probabilities for $d=5$, cycles$=54$, shots$=200$k. Stacked regions indicate component contributions (colors matching (\textbf{e})).
    Detectors are sorted by spatial coordinate with increasing probability and time-ordered within each.
    \textbf{e}: Distributions of detection contributions for each error type. CZ: controlled-Z, DD: dynamical decoupling, RO: readout, H: Hadamard, NL: nonlinear (the contribution beyond a linear budget approximation \suppsec{A4}), I: non-DD idle, LR: leakage removal, R: reset.
    }
    \label{fig:main_hexagonal}
\end{figure}

The assumed necessity of a square lattice for the surface code has been a driving influence on superconducting device layouts, with experiments to date implementing four local couplings on each qubit~\cite{krinner_realizing_2022, marques_logical-qubit_2022, google_quantum_ai_suppressing_2023, google_quantum_ai_quantum_2025}. 
This strong device requirement adds substantial design complexity and has limited some architectures from implementing the surface code~\cite{hertzberg_laser-annealing_2021, morvan_optimizing_2022, chamberland_topological_2020}.

Using a time-dynamic circuit featuring an alternative detecting region tiling, a recent proposal provides an embedding of the surface code on a hexagonal grid~\cite{mcewen_relaxing_2023}.
In \fig[a]{main_hexagonal} we highlight the key insights of this proposal. 
The first insight is that the last two-qubit gate layer of each QEC cycle can be altered to match the first layer with the CNOT orientation inverted, using the same qubit-qubit coupling layer twice in each cycle. 
This results in the expanding detecting region being shifted laterally while preserving the square weight-4 pattern of stabilizers. 
The second key insight is to use a time-reversed circuit every second cycle, refocusing the detecting region to its starting position during the contraction by an opposite lateral shift.
This pattern of two opposing detecting regions is sufficient to tile the space-time bulk of the circuit. We include details on the construction of detecting regions at the spatial boundaries in \suppsec{C1}.

We benchmark the hexagonal implementation using memory experiments run on a superconducting quantum processor. 
We first prepare the $X$ or $Z$ logical state, apply a variable number of error correction cycles, and finally measure the logical qubit in the initialized logical basis.
We use the same 72-qubit device as in Ref.~\cite{google_quantum_ai_quantum_2025}, which features flux-tunable capacitive couplers~\cite{foxen_qubit_2018, yan_tunable_2018}. 
We tune the unused couplers to their zero-coupling bias. 
The distance-5 hexagonal lattice circuit uses 49 computational qubits and 64 couplers, 20\% fewer than the 80 couplers used in the standard implementation. 
Even with fewer physical couplers used, the bulk detecting regions of both standard and hexagonal lattice implementations touch 22 CNOT gates, leading to similar sensitivity to two-qubit gate errors \suppsec{B}.
We optimize the operating frequencies of the gates taking into account the geometry of the circuit as in Ref.~\cite{klimov_optimizing_2024}. 
Additionally, we use the Data Qubit Leakage Removal (DQLR) technique developed in Ref.~\cite{miao_overcoming_2023} to mitigate the impact of leakage on the logical performance.

We run the memory experiment for a distance-5 hexagonal surface code and five tiling distance-3 codes.
In \fig[b]{main_hexagonal} we report the logical error probability at varying circuit depths, along with the logical error rate per cycle for different decoders (\fig[c]{main_hexagonal}), ordered from fastest to most accurate.
Using the Harmony decoder~\cite{shutty_efficient_2024} with a noise model prior trained using reinforcement learning~\cite{sivak_optimization_2024}, we measure a distance-5 logical error rate of $\epsilon_5 =0.270(3)\%$ and a mean distance-3 logical error rate of $\epsilon_3 =0.580(2)\%$ (averaged over five codes), fit with the same procedure described in \cite{google_quantum_ai_quantum_2025}. 
From this, we extract an error suppression factor of $\Lambda_{35, \text{hex}} = \epsilon_3 / \epsilon_5 = 2.15(2)$, indicating the surface code can be operated on a hexagonal lattice with performance matching the standard square lattice implementation ($\Lambda_{35, \text{standard}} = 2.14(2)$)~\cite{google_quantum_ai_quantum_2025}.

To build confidence in our results, we measure the fidelity of each component of the error correction circuit through independent benchmarking.
From these benchmarks and the circuit's detecting region description, we analytically compute a linear error budget for each region's detection probability, shown in \fig[d]{main_hexagonal} and accurate to first order in physical error probabilities \suppsec{A4}. 
We validate this budget by computing the root-mean-square (RMS) between the analytic prediction and measured value for detection probabilities, finding a RMS of $6.95\times 10^{-3}$ for the distance-5 code over 54 cycles \suppsec{C3}.
In \fig[e]{main_hexagonal}, we summarize the distribution of different physical error contributions to the detection probabilities.
The dominant source of errors are the entangling gate (CZ), the idling error on data qubits during measurement and reset (Dynamical Decoupling, or DD), and readout (RO). 
Since the boundary detectors use twice-fewer entangling gates, the CZ contribution is halved.
Single-qubit gates, leakage removal, and reset combined make up only 15\% of the detection budget, highlighting the quality of these operations.
\vspace{-3mm}
\section*{Walking Surface Code}
\vspace{-1mm}
\begin{figure*}
    \centering
    \includegraphics{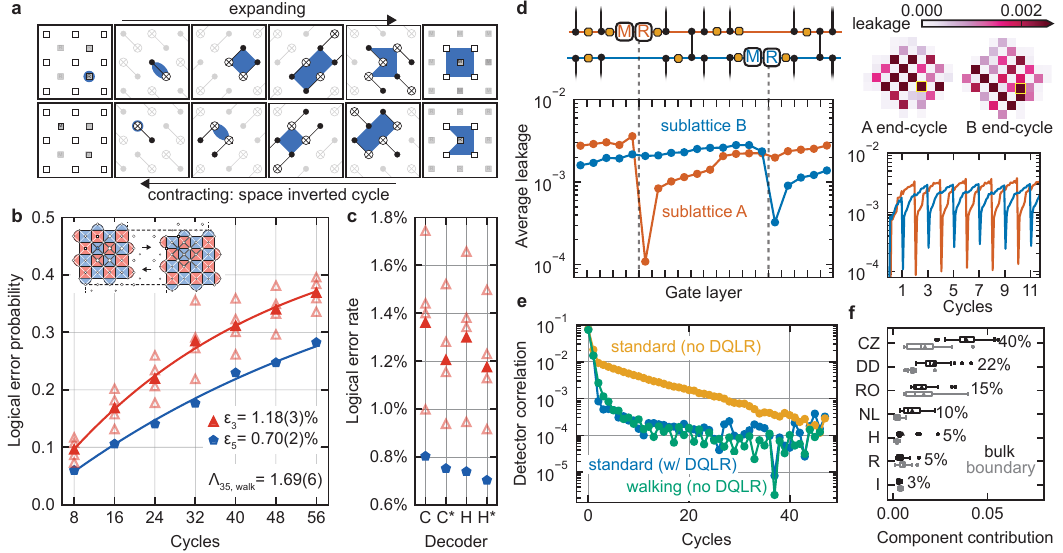}
    \caption{\textbf{Walking implementation.}
    \textbf{a}: Detecting regions slices for a Z-type bulk detector in the walking circuit. 
    \textbf{b}: Measured logical error probability for a distance-5 and embedded distance-3 surface codes implemented using the walking circuit and decoder H*.
    \textbf{c}: Comparison of different decoders. 
    \textbf{d}: Measured average leakage on the A and B data qubit sublattices after each walking circuit gate layer, with leakage data per qubit shown in the upper-right corner. To mitigate an outlier, the highlighted qubit used a single-level reset instead of a multi-level reset. Average leakage for both sublattices over 12 cycles is shown in the lower right. The steady state is reached rapidly, and no rise in leakage over time is visible.
    \textbf{e}: Comparison of the average detector autocorrelation for three circuits: the standard surface code without DQLR (orange), with DQLR (blue), and the walking circuit without DQLR (green). Long-time correlations in the standard code without DQLR are removed using walking or DQLR.
    \textbf{f}: Detector error budget from the component benchmarks.}
    \label{fig:main_walking}
\end{figure*}
In the standard circuit construction for a QEC code, a static set of data qubits are used to support logical information, and static measure qubits are added to perform the parity checks.
The strict distinction between these two categories has a direct consequence in hardware implementations. 
For example, the data qubits always hold important logical information and therefore are never replaced or reset to their ground state $\ket{0}$, allowing them to accumulate non-computational errors.
In superconducting circuits, leakage to $\ket{2}$ and higher represents one such accumulating error, limiting experimental performance~\cite{sundaresan_matching_2022, krinner_realizing_2022} and requiring hardware intervention~\cite{mcewen_removing_2021, miao_overcoming_2023, marques_all-microwave_2023, lacroix_fast_2023}.
In cold atom implementations, qubit loss when the atoms escape the trap is another critical example of an accumulating error~\cite{bluvstein_quantum_2022, chow_circuit-based_2024, radnaev_universal_2024, reichardt_logical_2024}. 
In the framework of detecting regions demonstrated here, the distinction between measure and data qubits is not necessary. 
It has been proposed that the surface code can be implemented using a circuit where each qubit alternates each cycle between the measure and data roles~\cite{mcewen_relaxing_2023, camps_leakage_2024}.
These circuits also shift the physical qubit support of the logical qubit and permit movement through space, so we refer to them as \emph{walking circuits}.

In a walking circuit, each physical qubit experiences a measurement and an opportunity for reset or replacement every two cycles.
For superconducting qubits, if a multi-level reset strategy is used at each cycle, the walking circuit presents the advantage of periodically removing leakage from all qubits.
For neutral atoms, it relaxes the necessary atom lifetime between opportunities for loss detection and replacement, limiting the errors induced by a loss event and allowing rapid atom replacement with a minimal impact on logical performance.

Walking is permitted by the gate layers and detecting region time-slices shown in \fig[a]{main_walking}. 
Using this walking circuit, we benchmark the scaling of a logical memory on a 105-qubit superconducting quantum processor~\cite{google_quantum_ai_quantum_2025} (a different device than was used for the hexagonal implementation), with logical error probability results shown in \fig[b]{main_walking} and logical error rate per cycle for different decoders in \fig[c]{main_walking}.
Here, we realize the distance-5 and embedded distance-3 walking circuits without any explicit leakage removal strategy other than multi-level reset (no DQLR).
In order to allow room for walking, the distance-5 circuit uses 58 qubits, more than the standard circuit's 49, leading to a slight overhead in logical error rate.
For a logical memory experiment, we measure distance-5 and average distanced-3 logical error rates of $\epsilon_3 = 1.18(3)\%$, $\epsilon_5 = 0.70(2)\%$, leading to an error suppression factor $\Lambda_{35, \text{walk}}=1.69(6)$.
The median CZ Pauli error of $3.56\times 10^{-3}$ used in the walking circuit is $47\%$ larger than the hexagonal implementation \suppsec{D2}, leading to a reduction in $\Lambda_{35}$.

To benchmark the walking circuit's leakage dynamics, we measure the leakage population at each time-slice of the circuit, separated into the two sub-grids of qubits as shown in \fig[d]{main_walking}. 
At each moment when a reset is applied to a group of qubits, the leakage population on that group drops by around an order in magnitude, followed by a ramp in population until they are reset again two cycles later. 
The staggered multi-level reset applications limit leakage population buildup over many cycles, preventing a rising contribution to the overall logical performance as the code continues.

Leakage also causes time-correlated errors that are not properly handled by the decoder. 
We directly probe these correlations by measuring the detector autocorrelation in \fig[e]{main_walking}, which is the average covariance of the detecting events separated by $n$ cycles.
While two-round correlations are expected from time-like Pauli errors (e.g. readout or reset errors), this measurement reveals longer time correlations induced by non-Pauli errors. 
Without leakage removal in the standard code, we measure a long correlation tail lasting nearly 50 cycles, while the walking circuit and the standard code with leakage removal significantly reduce these long-time correlations.
\begin{figure*}[ht]
    \centering
    \includegraphics{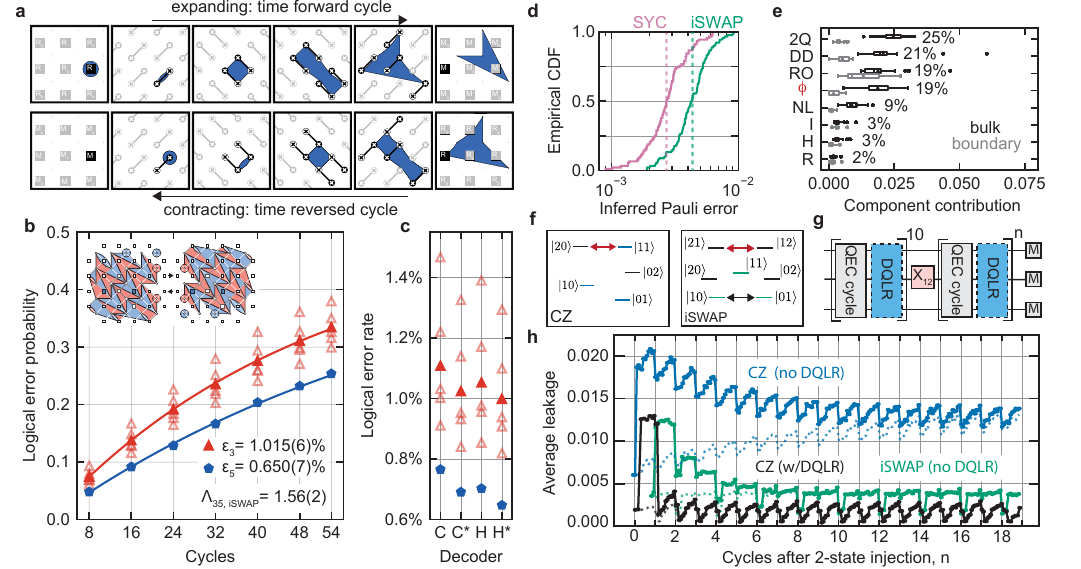}
    \caption{
    \textbf{iSWAP implementation.} 
    \textbf{a}: Detecting regions slices for a Z-type bulk detector in the iSWAP circuit, written in terms of  CX-SWAP gates, equivalent to the iSWAP gate under single qubit rotations \suppsec{E1}.
    \textbf{b}: Measured logical error probability for a distance-5 and embedded distance-3 surface codes implemented using the iSWAP circuit and decoder H* 
    \textbf{c}: Comparative decoder performance. 
    \textbf{d}: Measured Sycamore (pink) and iSWAP (green) inferred two-qubit Pauli errors from XEB, with the difference arising from the c-phase.
    \textbf{e}: Detector error budget from component benchmarks.
    \textbf{f}: Level diagram for the CZ and iSWAP gates (single- and two- excitation manifold shown), with computational states colored, and primary leakage transport highlighed (red arrows). 
    \textbf{g}: Leakage injection experiment with resulting data shown in (\textbf{h}). After 10 rounds of error correction, $\ket{2}$ state leakage is injected on a data qubit, and leakage is measured at each time-slice after. 
    \textbf{h}: Measured leakage slicing after injection for the standard CZ implementation with (black) and without (blue) DQLR and for the iSWAP implementation without DQLR (green). The reference experiments without injection are shown as dotted lines. 
    }
    \label{fig:main_iswap}
\end{figure*}
In addition to these leakage measurements, we apply the same detector budgeting techniques introduced in the previous section to reveal the contributions of each physical error mechanism on detection probabilities in the circuit (\fig[f]{main_walking}). 
Our Pauli model produces detections that have a weaker correlation with experiment (RMS $= 2.01 \times 10^{-2}$, see \suppsec{D3}) than the hexagonal implementation, indicating that some error mechanisms are not fully captured by the Pauli model, but reveals the same dominant contributions as the hexagonal implementation. 
The CZ contribution is larger as expected from the reduced gate performance. 
Even with the reduction in logical performance and $\Lambda_{35}$, our results demonstrate that a walking surface code can suppress errors as distance increases with a built-in leakage removal, paving the way for new architectures that can use this property to their advantage. 
\vspace{-3mm}
\section*{iSWAP Surface Code}
\vspace{-1mm}
The third implementation explored in this work uses the iSWAP family of gates for the entangling layers instead of the standard CNOT/CZ family. 
It was previously unclear how to implement quantum error correction using the iSWAP gate, as it distorts the stabilizers, preventing them from returning to their initial static code state. 
However, using the dynamic circuit technique of time-reversing every second cycle, we can find complementary distortions of the stabilizers that permit the iSWAP gate to preserve the structure of the surface code. 
In \fig[a]{main_iswap}, we show the expansion and contraction of a single detecting region. 
The end-cycles forms a distorted \emph{arrowhead} shape instead of the usual square. 
Nonetheless, the arrowhead pattern can be clearly identified as a two-colorable checkerboard of weight-4 X and Z stabilizers, sufficient for QEC of the surface code.

In \fig[b]{main_iswap} and \fig[c]{main_iswap}, we benchmark the logical error rate of a distance-5 and embedded distance-3 iSWAP implementation of surface codes on our 72-qubit device (the distance-5 circuit uses 57 qubits).
Using decoder H*, we measure logical error rates of $\epsilon_5 =0.650(7)\%$ and $\epsilon_3 =1.015(6)\%$ averaged over the five distance 3 codes, here again without using DQLR (relying on multi-level reset as the only active source of leakage removal). We extract the error suppression factor $\Lambda_{35, \text{iSWAP}} = 1.56(2)$, demonstrating the viability of a surface code implemented with iSWAP gates.

In previous experiments, the so-called \textit{Sycamore} gate using a short, on-resonance $\ket{01} - \ket{10}$ interaction has been used as an entangling resource to realize beyond-classical circuits \cite{arute_quantum_2019, morvan_phase_2024}.
The Sycamore gate can be decomposed as an iSWAP gate and a controlled-phase gate, with arbitrary angle $\phi$ \suppsec{E2}. 
While this controlled-phase is not a problem for NISQ experiments \cite{mi_information_2021}, it adds a Pauli error per gate of $\phi^2 / 16$ for stabilizer codes that require a strict Clifford gate, as confirmed by coherent simulations.  
In this work, we choose an iSWAP gate length balancing coherent error from the c-phase and incoherent error, arriving at a longer gate of order \SI{60}{ns}, roughly twice the CZ length.
This additional c-phase error channel can be reduced by tailoring the hardware to the iSWAP, for example by increasing the anharmonicity or using a different gate architecture \cite{sung_realization_2021}.  
For this gate length, we measure an average c-phase of $\phi=146~\text{mrad}$, yielding a Pauli error contribution to the gate of $1.33\times 10^{-3}$ as shown in \fig[d]{main_iswap}, for a median inferred iSWAP Pauli error of $4.28\times 10^{-3}$ \suppsec{E3}, $76\%$ larger the CZ gates used in the hexagonal implementation.
This c-phase error accounts for roughly 19\% of the detection probability in the bulk, as shown in \fig[e]{main_iswap}. 
Although our present device was optimized to realize CZ gates, the iSWAP c-phase error could be suppressed by increasing the anharmonicity in future designs. 

Despite the lower gate fidelity on current hardware, one appeal of using iSWAP instead of CZ is the leakage generation and transport. 
As shown in \fig[f]{main_iswap}, the iSWAP's on-resonance $\ket{10} - \ket{01}$ interaction is within the computational subspace and as a result does not directly populate non-computational states, minimizing leakage generation. 
When a higher excited state such as $\ket{2}$ is present, it will transport to other qubits through the resonant $\ket{21} - \ket{12}$ interaction, and this transport can aid in leakage removal by spreading leakage from data qubits to measure qubits where a multi-level reset gate is performed each cycle.
This behavior is in contrast to the CZ gate, which implements an on-resonance $\ket{20} - \ket{11}$ interaction with a precise timing calibrated to remove population from the $\ket{2}$ state. Small errors in CZ calibration can thus generate leakage outside the computational space. 

We probe the effect of leakage during memory experiments in \fig[h]{main_iswap} by measuring the population of excited states following the injection of a $\ket{2}$ state on the center-most bulk data qubit after 10 cycles as shown in \fig[g]{main_iswap}, here using the distance-5 code.
We repeat this experiment for three surface code implementations: the standard CZ implementation with and without DQLR, and the iSWAP implementation (which does not use DQLR).
In a reference experiment where no leakage is injected (dotted lines), the standard surface code without DQLR reaches a steady state with oscillating average leakage between $1.1\%$ and $1.4\%$ per qubit due to the large leakage generation of the CZ gates. With DQLR, the average leaked population is considerably reduced as expected, oscillting between $0.05\%$ and $0.3\%$. 
With the iSWAP implementation, the average leakage population reaches an oscillating steady state between $0.1\%$ and $0.45\%$, limiting leakage well below the standard implementation without DQLR, indicating the superior leakage properties of the iSWAP gate compared with the CZ gate.
Of all three, the standard CZ code with DQLR leads to the best leakage performance.

When the $\ket{2}$ state is injected after a reset moment, all three implementations spike in their leaked population. 
The standard implementation without DQLR eliminates the extra leaked population within 10 cycles, whereas the DQLR version eliminate the leaked state within 2 cycles and the iSWAP in 4 cycles. This demonstrates that the iSWAP implementation can mitigate leakage well without explicit leakage removal, and adding DQLR to the iSWAP implementation could further improve its leakage characteristics. 
\vspace{-3mm}
\section*{Scaling and outlook}
\vspace{-1mm}
\begin{figure}[tb!]
    \centering
    \includegraphics{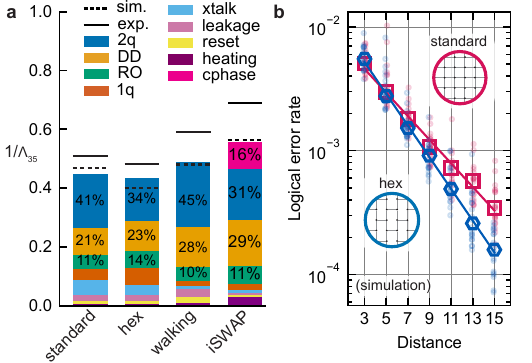}
    \caption{\textbf{Lambda budgeting and scaling.} \textbf{a}: $1/\Lambda_{35}$ error budgets for the hexagonal, walking, and iSWAP implementations, compared with the standard implementation from \cite{google_quantum_ai_quantum_2025}. The relative percentages of the dominant categories are given.
    \textbf{b}: Simulated logical error rate vs distance for hardware tailored to the hexagonal circuit (3 couplers per qubit) compared with a standard square grid layout with the same TLS landscape. Open squares (standard code) and hexagons (hex code) are error rates averaged over multiple simulations (circles).}
    \label{fig:main_scaling}
\end{figure}
To compare surface code implementations on current hardware, in \fig[a]{main_scaling} we show contributions of each error component to the inverse scaling factor, $1/\Lambda_{35}$, which can be expanded as a linear sum in physical error rates at first order \suppsec{B}. The total cycle times of the implementations are similar (hex: $\SI{1086}{ns}$, walking: $\SI{1028}{ns}$, iSWAP: $\SI{1064}{ns}$), allowing for direct comparison.
These simulations go beyond the Pauli models used to construct the detector error budgets by including effects such as leakage and crosstalk. 
For all codes, these leakage and crosstalk contributions are still small relative to the two qubit gates, dynamical decoupling (DD), and readout error. The cphase contribution for the iSWAP code is significant, accounting for $16\%$ of the budget, and causes the iSWAP code's simulated $1/\Lambda_{35}$ to be larger than the other three implementations.

For the hexagonal implementation, while the performance we measure already matches state of the art, tailoring the device to a honeycomb or brickwall lattice will reduce the number of connections needed from 2 to 1.5 per qubit (taking into account that qubits share connections).
In principle, it will also reduce the number of potential error channels by reducing the complexity of the device.
From a frequency optimization perspective, the lower connectivity reduces the number of variables, simplifying the optimization step. 
In \fig[b]{main_scaling}, we simulate the potential gain of tailoring the hardware to a hexagonal lattice by accounting for realistic device inhomogeneities in hardware parameters as well as TLS density with a frequency optimization step \cite{klimov_optimizing_2024} \suppsec{G}. 
From these simulations we predict that simplifying the hardware by removing one coupler per qubit can lead to a $15\%$ improvement of the scaling factor $\Lambda$.
Such an increase in $\Lambda$ compounds favorably with a reduction in number of couplers needed, simplifying the design and number of control lines.

With these data, we have demonstrated state-of-the-art error suppression from distance-3 to distance-5 for the hexagonal, walking, and iSWAP surface code implementations.
This proves that dynamic circuits are a viable approach to fault-tolerance, while also unlocking the ability to tailor the error correction circuit to hardware, presenting new opportunities for coupler design, multiplexing, and leakage removal intrinsic to the operation of QEC itself.
In addition, our budgeting and leakage results provide direct insight into how using dynamic circuits uncovers new tradeoffs for realizing quantum error correction. 
We envision that other unexplored avenues for dynamic circuits may present further benefits, and
our experiments solidify the prospects for the co-design of QEC and hardware as the state of the art for quantum error correction at scale.

\vspace{-3mm}
\section*{Contributions}
\vspace{-1mm}
A. Morvan, M. McEwen, and A. Eickbusch conceived and led the project. 
A. Morvan and A. Eickbusch performed the calibrations and measurements.
A. Morvan, A. Eickbusch and M. McEwen wrote the manuscript with input from all the authors. 
A. Morvan, A. Eickbusch, V. Sivak, M. McEwen, J. Atalaya and J. Claes wrote the supplements. 
M. McEwen, A. Eickbusch, A. Morvan, D. Bacon and C. Gidney designed the error correction circuits for the experiments.
V. Sivak performed the reinforced learning decoding on all datasets and performed the simulation with gate frequency optimization and scaling to larger distance. 
A. Bourassa developed coherent error corrections for dynamic circuits.
J. Atalaya, J. Claes, D. Kafri, A. Eickbusch, and M. McEwen performed the Pauli and hardware-accurate simulations. 
A. Eickbusch and A. Morvan developed the error budgeting for detectors.
Z. Chen, A. Bengtsson, and A. Green performed initial bringup and calibration of the device.
P. Klimov, W. Livingston, A. Pizzuto, and V. Sivak developed the algorithm specific frequency optimizations for these experiments. 
A. Bourassa, G. Roberts, K. Satzinger, and M. Neeley developed the infrastructure to translate QEC circuits to the hardware.
A. Opremcak and A. Bengtsson performed calibration of the readout and dynamical decoupling.
K. Miao and N. Zobrist performed the calibration of the reset and Leakage removal operations. 
C. Warren, and J. Gross developed the c-phase characterization of the iSWAP gate.
All authors contributed to building the hardware and software, and writing the manuscript.
\vspace{-3mm}
\section*{Acknowledgements}
\vspace{-1mm}
We thank Michel Devoret for extensive discussion on the manuscript.
\vspace{-3mm}
\section*{Corresponding authors}
\vspace{-1mm}
Correspondence and requests for materials should be addressed to Alexis Morvan (amorvan@google.com), Alec Eickbusch (aleceickbusch@google.com) and Matt McEwen (mmcewen@google.com).
\vspace{-3mm}
\section*{Competing Interests}
\vspace{-1mm}
The authors declare no competing interests.
\vspace{-3mm}
\section*{Code and data availability}\label{supp}
\vspace{-1mm}
The datasets generated and analysed for this study are available at \href{https://doi.org/10.5281/zenodo.14238907}{https://doi.org/10.5281/zenodo.14238907}.

\newpage
\onecolumngrid

\vspace{1em}
\begin{flushleft}
{\small \textbf{Google Quantum AI and Collaborators}}

\bigskip
{\small
\renewcommand{\author}[2]{#1$^\textrm{\scriptsize #2}$}
\renewcommand{\affiliation}[2]{$^\textrm{\scriptsize #1}$ #2 \\}

\newcommand{\corrauthora}[2]{#1$^{\textrm{\scriptsize #2}, \ddagger}$}
\newcommand{\corrauthorb}[2]{#1$^{\textrm{\scriptsize #2}, \mathsection}$}

\newcommand{\xGoogle}{\affiliation{1}{Google Quantum AI, Santa Barbara, CA 93117, USA}}

\newcommand{\xUCONN}{\affiliation{2}{Department of Physics, University of Connecticut, Storrs, CT, USA}}

\newcommand{\xUMass}{\affiliation{3}{Department of Electrical and Computer Engineering, University of Massachusetts, Amherst, MA}}

\newcommand{\xAU}{\affiliation{4}{Department of Electrical and Computer Engineering, Auburn University, Auburn, AL, USA}}

\newcommand{\Google}{1}
\newcommand{\UCONN}{2}
\newcommand{\UMass}{3}
\newcommand{\AU}{4}

\corrauthora{Alec Eickbusch}{\Google},
\corrauthora{Matt McEwen}{\Google},
\author{Volodymyr Sivak}{\Google},
\author{Alexandre Bourassa}{\Google},
\author{Juan Atalaya}{\Google},
\author{Jahan Claes}{\Google},
\author{Dvir Kafri}{\Google},
\author{Craig Gidney}{\Google},
\author{Christopher W.~Warren}{\Google},
\author{Jonathan Gross}{\Google},
\author{Alex Opremcak}{\Google},
\author{Nicholas Zobrist}{\Google},
% ...
\corrauthora{Alexis Morvan}{\Google}.

\bigskip

\xGoogle
\xUCONN
\xUMass
\xAU

{${}^\ddagger$ These authors contributed equally to this work.}

}
\end{flushleft}

\twocolumngrid

\break
\newpage
\bibliography{bib.bib}
    
\end{document}


\title{Supplementary material for ``Demonstrating dynamic surface codes''}
    \date{\today}
    \author{Google Quantum AI and Collaborators}
    
    \maketitle
    
    \newpage
    \onecolumngrid
    \appendix
    \renewcommand\thefigure{\thesection.\arabic{figure}} 
    \renewcommand\thetable{\thesection.\arabic{table}} 
    
    \setcounter{figure}{0}  
    \setcounter{table}{0}
    
\title{Supplement to Demonstrating dynamic surface codes}

\maketitle

\tableofcontents
    \newpage
\section{Statistical analysis of QEC experiments}
\label{sec:supp_stats}
In this appendix, we provide statistical methods and error budget modeling. We first establish a framework for analyzing detection events, including the average and covariance. We link these statistical measures to underlying Pauli error channels. We then explain how to construct both linear and exact error budgets analytically within a Pauli model, enabling the assessment of individual error mechanism contributions. Then, we outline experimental benchmarking techniques for characterizing the various quantum gates and process errors in QEC experiments.

\subsection{Detector probability}

A single QEC memory experiment returns a dataset composed of a bits that form the \textit{detection events}, $d_i$, where each entry correspond to a detector in the circuit. It also return a final \textit{measured logical observable} bit that we will denote as $L$. The set of detection events $\{ d_i\}_{i \in \text{detectors}}$ and the logical observable $L$ is passed into the decoder, along with an error model, to return the corrected logical observable $L_c$. In the following, we will analyze these detection events to describe our experiments and validate our models.
We perform several repetitions, $N_s$, of the same experiment in order to gather statistical confidence into our logical error rate. We also vary the initial data-qubit states, specified by $N_b$ different random bitstrings, to mitigate bias. In the following we will denote the expectation value of a detector or a logical observable averaged over both repetitions and initializations $N_s \times N_b$ as:
\begin{equation}
    \E{X} = \frac{1}{N_s N_b} \sum_{i \in \text{init.}} \sum_{r \in \text{rep.}} X_{i,r}.
\end{equation}
Hence the probability of a detector $d_i$ firing is written as $\E{d_i}$.

We will describe errors as Pauli channels $\mathcal{E}$ following stim \cite{gidney_stim_2021} conventions. We define a Pauli channel $\mathcal{E}$ as 
\begin{equation}
    \mathcal{E}(\rho) = (1-p) I + \sum_{P\in\mathcal{P} \backslash {I}} p(P) P \rho P
\end{equation}
where $\rho$ is the density matrix, $I$ the identity and $\mathcal{P}$ the Pauli group, such that $p(P)$ is the probability associated with a $P$ error. To simplify discussion in the rest of the supplement, we will denote $\mathcal{E}_{P}$ as a channel associated with a single Pauli operation,
\begin{equation}
    \mathcal{E}_{P}(\rho) = (1-p) I + p P \rho P.
\end{equation}

We will use depolarization noise for conciseness, but note that the probability of each Pauli will be considered independent. The single-qubit depolarization channel is given as
\begin{equation}
    \mathcal{E}_{D1}(\rho) = (1-p) I + \frac{p}{3} X +\frac{p}{3} Y + \frac{p}{3} Z
\end{equation}
and the two-qubit depolarization reads:
\begin{equation}
    \mathcal{E}_{D2}(\rho) = (1-p) I + \sum_{P\in\mathcal{P}_2 \backslash {I}} \frac{p}{15} P \rho P.
\end{equation}

\begin{figure}[h]
    \centering
    \includegraphics{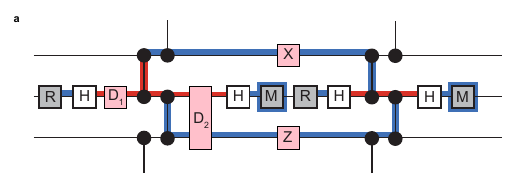}
    \caption{\textbf{Detector with errors}: a single detector of the repetition code with different type of errors. The error is indicated as red boxes. "D" indicates depolarization noise, $\text{D}_1$ for single qubit depolarization noise and $\text{D}_2$ for 2-qubit depolarization noise. In this case, the Z error commutes with the region of detector shown (blue=Z), so it does not trigger this detector.}
    \label{fig:supp_stats_detector_errors}
\end{figure}

To predict the detection probabilities from an error model, we constrain ourselves to a stochastic Pauli error model as described in \ref{fig:supp_stats_detector_errors} where each gate can be associated with an error channel with probability $p$. To compute the detection probability, we add through a probabilistic XOR the probability of each error channel $\mathcal{E}$ that anti-commutes with the detecting region. 
The probabilistic XOR for two stochastic Pauli channels $\mathcal{E}_1$ with probability $p_1$ and $\mathcal{E}_2$ with probability $p_2$ reads:
\begin{equation}
    p(\mathcal{E}_1 \oplus \mathcal{E}_2) = p_1 (1-p_2) + p_2 (1-p_1)
    \label{eq:xor}
\end{equation}

This formula accounts for the fact that two errors will not trigger a detector. In the limit of very small probability, this formula can be approximated as a linear sum. This can be recast into the following equation, which is just a transformation from boolean variable ($0, 1$) to Spin 1/2 $(-1, 1)$:
\begin{equation}
    1 - 2p(\mathcal{E}_1 \oplus \mathcal{E}_2) = (1 - 2 p_1) (1 - 2p_2).
\end{equation}
This allows the generalization to arbitrary number of errors channels, and link them to the expected value of a detector $d_n$:
\begin{equation}
    1 - 2\E{d_n} = 1 - 2p \left( \bigoplus_{i \in \mathcal{A}_n} \mathcal{E}_i \right) = \prod_{i \in \mathcal{A}}(1-2p_i)
    \label{eq:det expect}
\end{equation}
where now, $\mathcal{A}_n$ is the set of errors channels that anti-commute with the detecting region of the detector $d_n$.
By taking the log, each contribution (different error channels) can be added independently:
\begin{equation}
    \ln\left(1 - 2 \text{E}[d_n] \right) = \sum_{i \in \mathcal{A}_n} \ln\left(1 - 2 p_i \right)
    \label{eq:ln det expect}
\end{equation}
valid for detection expectation values $\text{E}[d_n] < 0.5$.

\subsection{Detector probability sampling noise}
Detection probabilities are the average of detection events for a single detector $d_i$ over repetitions and initialization bitstrings $\E{d_i}$. If we suppose that all errors follow a binomial distribution, the variance for the detection probability is given by:
\begin{equation}
    \text{Var}[d_i] =  \text{E}[d_i] \left(1- \text{E}[d_i]\right)
\end{equation}
and the sampling noise or shot noise is given by $\sigma_{i} = \text{std}[d_i]/ \sqrt{N} =  \sqrt{\text{Var}[d_i] / N}$ which has the same unit as the average detection probability.

\subsection{Covariance of two detectors}
Intuitively, the covariance of two detectors gives insight into the errors that trigger both detectors. The probability of triggering one detector or the other is given by 

\begin{equation}
    2 \E{d_n d_m} = \left(\E{d_n} + \E{d_m} - p \left( \bigoplus_{i \in \mathcal{A}_n \cap \mathcal{A}_m} \mathcal{E}_i \right) \right)
\end{equation}
where the last term is an xor over the channels that trigger both detectors. This last term is the one we are interested in experimentally, since it yields more localized information on the circuit error. A proxy for this quantity is the covariance of the detectors that can be written as 
\begin{equation}
    C_{nm} = \E{d_n d_m} - \E{d_n}\E{d_m} = \E{(d_n - \E{d_n}) (d_m - \E{d_m})}.
\end{equation}
The sampling noise for the covariance is given by
\begin{align}
\text{Var}[C_{nm}] &= \E{(d_n- \E{d_n})^2 (d_m- \E{d_m})^2}  - C_{nm}^2 \\
&= C_{nm} \left(1-2\E{d_n}\right) \left(1-2\E{d_m}\right) \\
& + \E{d_n} \E{d_m}\left(1-\E{d_n}\right) \left(1-\E{d_m}\right)
\end{align}

\subsection{Error budgeting for detections}

From a stochastic Pauli noise model, we can construct an error budget for detectors analytically. 
The method presented here could be used to find the contribution of any individual error mechanism in the noise model to the detection probability of any detector in the circuit. 
Here, we focus on finding the error contribution for different groupings of errors (such as two-qubit gate errors, readout errors, etc.) 

\subsubsection{Linear error budget}
To build a linear budget, we start with \eq{det expect} and expand to first order in error probability to find
\begin{equation}
    \E{d_n} = \sum_{i \in \mathcal{A}_n} p_i + \mathcal{O}(p_i p_j).
\end{equation}

We then group all channels into non-overlapping sets $\left\{a_j\right\}$. For example, $a_0 = \left\{\mathcal{E}_k \text{ s.t. } \mathcal{E}_k \text{ is a single-qubit gate error}\right\}$, $a_1 = \left\{\mathcal{E}_k \text{ s.t. } \mathcal{E}_k \text{ is a two-qubit gate error}\right\}$, etc. 

For each detector, we partition the set of anti-commuting errors according to these groupings of channels,
\begin{align}
    &\E{d_n} = \sum_{a_j} \sum_{i \in \mathcal{A}_n \cap a_j} p_i +  \mathcal{O}(p_i p_j).
\end{align}

The linear contribution $C_{n,j}$ of error set $a_j$ to the detection probability $\E{d_n}$ is then

\begin{align}
    C_{n,j} = \E{d_n} -   \E{d_n} \Big\rvert_{a_j = 0}
    \label{eq:lin det contribution}
\end{align}
where $\E{d_n} \Big\rvert_{a_j = 0}$ is the expectation value of detector $d_n$ with all errors in group $a_j$ turned off, and expectation values in \eq{lin det contribution} are calculated using \eq{det expect}.

By calculating the linear contribution using a subtraction near the operating point, many of the nonlinear terms cancel in the subtraction. 
As a result, this linearization can be more accurate than calculating the linear contribution of group $j$ by turning off all errors except group $j$.

The nonlinear (N.L.) contribution of $\E{d_n}$ is then the difference between the model expectation value and the sum total of all linear contributions, 

\begin{align}
    C_\text{n, \text{N.L.}} = \E{d_n} -   \sum_{a_j} C_{n,j}  \propto \mathcal{O}\left(p_i p_j\right),
    \label{eq:nonlin det contribution}
\end{align}
which increases with the detection fraction. 

\subsubsection{Exact error budget}

If we instead construct a budget starting from \eq{ln det expect}, the various error mechanisms can be disentangled, and nonlinear cross terms removed. 
In particular, we can analytically construct a budget using
\begin{equation}
    -\frac{1}{2}\ln\left(1 - 2 \E{d_n} \right) = -\frac{1}{2} \sum_{a_j} \sum_{i \in \mathcal{A}_n \cap a_j} \ln\left(1 - 2 p_i \right),
    \label{eq:ln det budget}
\end{equation}

where the disentangled contribution $\tilde{C}_{n,j}$ of error set $a_j$ to the quantity $-\frac{1}{2}\ln\left(1 - 2 \E{d_n} \right)$ is then

\begin{align}
    \tilde{C}_{n,j} = \left(-\frac{1}{2}\ln\left(1 - 2 \E{d_n} \right)\right) -   \left(-\frac{1}{2}\ln\left(1 - 2 \E{d_n} \Big\rvert_{a_j = 0} \right)\right).
    \label{eq:lin det contribution}
\end{align}

The factors $-1/2$ are included so after expending to first order in $p_i$ the linear budget is recovered.

In the main text figures, we use the linear budget, since it is easier to comprehend. 
The exact budget here is also useful, in particular as a verification step to ensure all error mechanisms are accounted for in the budget, since $\left(-\frac{1}{2}\ln\left(1 - 2 \E{d_n} \right)\right) - \sum_{a_j} \tilde{C}_{n,j} = 0$ should always hold. 
We have verified this is the case for all detector budgets presented in this work, justifying the use of a non-linear component contribution (NL) calculated as the difference between the simulated detection probability and the sum of all contributions, as explained in the linear error budget section above.

All calculations for analytic detections, covariances, and budgets are done using stim's detector error model feature (no sampling is involved). Component budgets for all detectors in a QEC circuit can be obtained within seconds once benchmark data for the Pauli model is provided.

\subsection{Experimental benchmarking and error model construction}

Here we describe how device errors are benchmarked and how the Pauli noise model used for detector budgeting and calibration validation is constructed.

\vspace{0.5cm}

\textbf{Single qubit gates:}
We measure the single qubit depolarization rate through parallel single qubit Randomized Benchmarking (RB)\cite{emerson_scalable_2005} on all the qubits.
RB data is usually reported as an \textit{average errors} $\varepsilon_{\text{RB}}$. 

In the Pauli noise mode, we translate the average RB error into a Pauli error rate for a single qubit depolarizing channel using:
\begin{equation}
    p_{1q} = \frac{3}{2}\varepsilon_{\text{RB}}.
\end{equation}

Since we use a depolarization model, this noise channel can be applied before or after the single qubits gates. We choose to follow the convention of applying them after.

\vspace{0.5cm}

\textbf{Measurement:}
For the measurement errors, we randomly prepare $\ket{0}$ or $\ket{1}$ and measure, using the same layers of parallel measurement as in the circuit. 
The measurement error for the Pauli noise model is then $p_m = P((0|1) + P(1|0))/2$, where $P(a|b)$ is the probability of measurement outcome $a$ when preparing state $\ket{b}$.
To model measurement error, the channel
\begin{equation}
    \mathcal{E}_{M}(\rho) = (1-p_m) I + p_m X \rho X
\end{equation}
is added \textit{before} each measurement gate, using the measured $p_m$ for each qubit in the corresponding measurement layer.

\vspace{0.5cm}

\textbf{Reset:}
To measure reset errors, we prepare all qubits in a reset layer in $\ket{0}$ or $\ket{1}$, apply the parallel reset operation, then measure.
The reset error is then $p_r = \left(P(1|0) + P(1|1)\right)/2$ where $P(a|b)$ is the probability of measurement outcome $a$ after reset when preparing state $\ket{b}$. 
To model reset error, the channel
\begin{equation}
    \mathcal{E}_{R}(\rho) = (1-p_r) I + p_r X \rho X
\end{equation}
is added after each reset gate, using the measured $p_r$ for each qubit in the corresponding reset layer.
\vspace{0.5cm}

\textbf{Two qubit gates:}
We measure the two qubit gate errors using parallel two-qubit cross-entropy benchmarking (XEB)\cite{arute_quantum_2019}.
The two-qubit gates are measured in the same parallel layers that are used in the QEC circuit for each implementation. 
The XEB is error usually reported as the average error $\varepsilon_{\text{XEB}}$. 
To find the two-qubit gate error (sometimes called the inferred error), we subtract the single qubit errors found from RB.

We translate the inferred gate error to a Pauli error rate for a 2-qubit depoliarzion channel using
\begin{equation}
    p_{2q} = \frac{5}{4}\varepsilon_{\text{XEB}}
\end{equation}
Since we use a depolarization model, the error can be applied before of after the gate. 
By convention we choose to apply it after the gate.

\vspace{0.5cm}

\textbf{Special case of the C-phase for the iSWAP gate:}
As described later in the main text, the iSWAP gate has a coherent cphase error component. 
Using coherent simulations, we have verified that this error is Pauli-twirled and can be modeled as a Pauli ZZ error in simulations.
For this error, we add an extra Pauli channel after the iSWAP gate as:

\begin{equation}
    \mathcal{E}_{\phi}(\rho) = (1-p_\phi) I + p_\phi ZZ \rho ZZ \quad \text{with} \quad  p_{\phi} = \frac{1}{16} \phi^2
\end{equation}
where $\phi$ is the measured cphase per two qubit gate, measured in parallel using the same gate layers as the circuit. 
To measure the cphase, we use the technique developed in \cite{gross_characterizing_2024} for measuring two-qubit gate coherent errors.

\vspace{0.5cm}

\textbf{Dynamical Decoupling:}
In our QEC circuits, dynamical decoupling is performed on data qubits during measurement and reset layers.
To benchmark the error associated with this operation, we initialize each data qubit in $\ket{1}$ or $\ket{+}$, perform $N$ rounds of dynamical decoupling in the same parallel layers used in the QEC circuit, and finally measure the data qubit purity with tomography. 
During the benchmark, measurement and reset is performed on the measure qubits during each DD repetition, matching the context of the QEC experiment.
The purity decay during DD is fit to an exponential, $A e^{-\epsilon n} \approx 1 - \epsilon n$ (approximation valid at small error) where $n$ is the number of DD cycles. The per-cycle purity decay error $\epsilon_1$ or $\epsilon_2$ when starting in $\ket{1}$ or $\ket{+}$ respectively is extracted from this fit. 

To add this error into our model, we estimate the decay during DD as a $T_1$ and $T_2$ channel, given by
\begin{equation}
    \mathcal{E}[\rho] = (1-\frac{\epsilon_1}{4} - \frac{\epsilon_2}{2}) I\rho I + \frac{\epsilon_1}{4} \left(X \rho X  + Y \rho Y \right) + \left(\frac{\epsilon_2}{2} - \frac{\epsilon_1}{4}\right) \left( Z \rho Z \right).
\end{equation}

In the iSWAP benchmark data, only $\epsilon_2$ is measured for historical reasons. 
For the iSWAP Pauli model, we assume $\epsilon_1 = t/T_1$, where $T_1$ is the qubit's $T_1$ and $t$ is the length of DD.

\vspace{0.5cm}

\textbf{Data Qubit Leakage Removal:}
In our implementations, Data Qubit Leakage Removal (DQLR)\cite{miao_overcoming_2023} is included in the hexagonal lattice circuit, but not in the walking circuit or iSWAP circuit. 
The error on data qubits during DQLR is measured by preparing the data qubits in either $\ket{1}$ or $\ket{+}$, applying repeated applications of DQLR in the same layers as the experiment, and performing qubit tomography. 
In this benchmark, an echoing sequence is used to remove the effect of background quasistatic phase error.
We also perform an identical control experiment, where no DQLR is performed. 
We compare the DQLR experiment with the control experiment to find the excess Pauli error on data qubits during DQLR. 
This excess error is added as a depolarizing error after DQLR gates in the Pauli noise model, in addition to the $T_1$ and $T_2$ channel calculated from measured coherences at idle and the DQLR timing.

\vspace{0.5cm}

We summarize all components of the Pauli model along with device benchmarks in the following table:
\begin{table}[h]
\centering
\begin{tabular}{|l|l|l|l|l|}
\hline
Gate         & Benchmark                                                    & Channel & Pos. & Parameters                                                               \\
\hline
single qubit & randomized benchmarking (RB) \cite{emerson_scalable_2005} & single-qubit dep.    & after        & $ p = 3/2\varepsilon_{\text{RB}}$ \\
measurement  & prep. random 0/1 + measure & bit-flip    & before       &   $p=\left(P(1|0) + P(0|1)\right)/2$  \\
reset        & prep. 0/1 + reset + measure & bit-flip   & after       &   $p = \left(P(1|0) + P(1|1)\right)/2$  \\
two qubit & XEB \cite{arute_quantum_2019} \& RB (inferred error used) & two-qubit dep.    & after        & $ p = 5/4 \varepsilon_{\text{XEB}}$ \\
iSWAP cphase & MEADD  \cite{gross_characterizing_2024} & $p_\phi ZZ \rho ZZ$    & after        & $ p_\phi = \phi^2/16 $ \\
DD & 
prep. $\ket{1}$ ($\ket{+}$), DD purity decay $\epsilon_1$ ($\epsilon_2$) & $ p_1 \left(X \rho X  + Y \rho Y \right) + p_2 Z \rho Z$ & after        & $p_1 = \epsilon_1/4,\, p_2 = \epsilon_2/2 - \epsilon_1/4$ \\
DQLR \cite{miao_overcoming_2023} & excess Pauli error during DQLR $\epsilon_e$ & single-qubit dep., bit-flip, phase-flip & after        & $p = \epsilon_e$, $T_1$, $T_2$\\ 
other idle & $T_1$ and $T_2$ measurement & bit-flip, phase-flip & after & $T_1$, $T_2$ \\
             \hline
\end{tabular}
\caption{Device benchmarks entering into the stochastic Pauli noise model used for rapid detection budgeting and verification. 
}
\end{table}

    \newpage
\section{Surface code simulation and $1/\Lambda$ error budgets}
\label{sec:Lambda-budget} 

\subsection{Simulation details}
\label{sec:sim_details} 

In addition to the Pauli noise model described previously, we also used the Pauli+ simulator 
to conduct more physically detailed simulations of each experiment. Since most of the simulation details are described in previous works ~\cite{google_quantum_ai_suppressing_2023, google_quantum_ai_quantum_2025}, we give a brief summary of the included error mechanisms. The simulations are based on \texttt{Cirq} Circuits representing each ideal experiment, but with additional quantum channels representing noise. The following noise mechanisms are included:
\begin{itemize}
    \item Decoherence mechanisms, as represented by $T_1$ decay, white noise dephasing ($T_{\phi}$), and passive heating from state $\ket{1}$ to state $\ket{2}$. 
    \item Readout and reset errors
    \item Excess leakage $\ket{11}\rightarrow \ket{02}$ (to higher frequency qubit) during two-qubit CZ gates. 
    \item Stray coupling crosstalk errors (due to unwanted nearest neighbor and diagonal neighbor interactions) during CZ gates.
    \item Unwanted CPhase errors during iSWAP gates (detailed below).
    \item Depolarizing error on data qubits during idling (measurement + reset), added to account for excess errors beyond what is predicted by decoherence.
    \item Depolarizing error after single-qubit gates, in excess of decoherence.
    \item Depolarizing error after two-qubit gates, in excess of decoherence, stray coupling crosstalk, excess leakage, and CPhase error.
\end{itemize}
For each simulation, the relevant noise parameters were extracted from experimental characterizations conducted at time at which data is taken. 

Specific component errors and relative sensitivities are detailed in Tables~\ref{Suppl:hex-error-budget}-\ref{Suppl:iswap-error-budget}. For each component error, the noted value $p_{\text{expt}}^{(c)}$ represents a dimensionless error rate associated with the operation. These error rates are generally expressed as the \emph{Pauli error} associated with a given operation. One exception is leakage heating, which is expressed as the Pauli error due to leakage accumulating over one round of error correction. This is estimated as $\frac{1}{2}\Gamma_{1\rightarrow 2} t_{r}$, where $\Gamma_{1\rightarrow 2} = 1 / (2 \textrm{ ms})$ is the rate at which state $\ket{1}$ heats to state $\ket{2}$ and $t_{r}$ is the duration of a single round of error correction. The single novel error mechanism introduced in this study was unwanted CPhase error during iSWAP gates. This is modeled as an additional unitary error $\exp(i\phi Z\otimes Z /4)$ applied after each iSWAP gate, with corresponding Pauli error $\sin^2(\phi/4)$.  

We construct the $1/\Lambda_{35}$ error budget for each of the dynamic surface codes following the methodology of \cite{google_quantum_ai_suppressing_2023,google_quantum_ai_quantum_2025}. We write $1/\Lambda_{35}$ as a sum of contributions from each error channel $c$ that is considered in the simulation:
\begin{align}
\frac{1}{\Lambda_{35}} = \sum_c w_c \, p_{\rm expt}^{(c)}
\label{eq:1_Lambda_budget_formula}.
\end{align}
The weights $w_c$ are the sensitivities of $\Lambda_{35}^{-1}$ to a uniform increase, $\delta p_c$, of the error probabilities of all components associated to the $c$th error channel from their error operation point values. From numerical simulations, we find a linear dependence between $1/\Lambda_{35}$ and $\delta p_c$, and then $w_c$ is obtained from a linear fit. In Eq.~\eqref{eq:1_Lambda_budget_formula}, $p_{\rm expt}^{(c)}$ is the mean of the error probabilities of all components associated to the $c$th error channel; their values are given in Tables~\ref{Suppl:hex-error-budget}--\ref{Suppl:iswap-error-budget}.  

\subsection{Error budget for hexagonal lattice surface code} 
\label{sec: hex-budget}
The sensitivities and the (measured) error operating point are respectively specified by the parameters $w_c$ and $p_{\rm expt}^{(c)}$ of  Table~\ref{Suppl:hex-error-budget} for the $c$th error channel, and the product $w_cp_{\rm expt}^{(c)}$ gives its contribution to $\Lambda_{35}^{-1}$. We find that errors related to CZ gates (first three rows of Table~\ref{Suppl:hex-error-budget}) contribute roughly 46\% of the $1/\Lambda_{35}$ budget. The next important contributor is data qubit idle errors (23\%), followed by readout/reset errors (17\%) and single-qubit errors (13\%). These relative contributions are similar to those obtained for the conventional square lattice surface code~\cite{google_quantum_ai_quantum_2025}. In the calculation of the error budget, we assume 95\% (15\%) fidelity of data qubit leakage removal (DQLR) for the leakage state $|2\rangle$ ($|3\rangle$). The fact that we use much smaller reset fidelity for state $|3\rangle$ is because, in experiment, DQLR is mainly tailored for state $|2\rangle$. Decoding was carried out using sparse blossom decoder~\cite{higgott_sparse_2023}.   

The predicted value of $\Lambda_{35}$ from a linear combination of the component errors
 and sensitivities, following Eq.~\eqref{eq:1_Lambda_budget_formula} and Table~\ref{Suppl:hex-error-budget}, yields $\Lambda^{\rm linear}_{35}= 2.3$ while direct calculation of $\Lambda^{\rm Pauli+}_{35}={\rm LER}(d=3)/{\rm LER}(d=5)$, using Pauli+ simulator of Section~\ref{sec:sim_details} to compute the logical error rates (LERs) for code distances 3 and 5, yields $\Lambda^{\rm Pauli+}_{35}= 2.5$. We note that these predicted values of $\Lambda_{35}$ are better than the experimental values, which are $\Lambda_{35}^{\rm expt,\, sparse-blosssom}=2.1$ and $\Lambda_{35}^{\rm expt,\, harmony}=2.14$ (mentioned in the main text) obtained using sparse blossom and harmony  decoders. This indicates that there are additional error mechanisms in  the experiment not included in the simulations.

\subsection{Error budget for walking surface code} 
\label{sec: walking-budget}

We specify similar sensitivities and measured error operating points for the walking surface code in Table~\ref{Suppl:walking-error-budget}. In this simulation, there is no DQLR, as the walking surface code intrinsically removes leakage. Notably, despite the absence of DQLR, the $7\%$ contribution to $\Lambda_{35}^{-1}$ from leakage is similar to the $6\%$ contribution in the hexagonal lattice surface code above, as well as the conventional surface code~\cite{google_quantum_ai_quantum_2025}, indicating the walking surface code is successfully removing leakage.

The predicted value of $\Lambda_{35}$ using the linear approximation of Eq.~\eqref{eq:1_Lambda_budget_formula} and the data in Table~\ref{Suppl:walking-error-budget} is $\Lambda_{35}^\text{linear}=2.12$, while directly evaluating $\Lambda_{35}$ from the Pauli+ simulator gives $\Lambda_{35}^{\text{Pauli}+}=2.27$. Both of these values are notably higher than the measured $\Lambda_{35}^\text{expt, harmony}=1.67$ given in the main text, indicating that we are likely missing some significant source of error in the walking code.

\subsection{Error budget for iSWAP surface code} 
\label{sec: iswap-budget}

We specify similar sensitivities and measured error operating points for the iSWAP surface code in Table~\ref{Suppl:iswap-error-budget}. In these simulations, we do not include DQLR as in the experiment. As the iSWAP gate induces much less leakage than a CZ gate, we find that  iSWAP leakage contributes much less to the overall $\Lambda_{35}^{-1}$ (roughly 1\% compared to 5\% for hexagonal and 6\% for walking surface codes). We also note that leakage heating has a more detrimental effect on this surface code due to lack of DQLR that leads to a leakage lifetime of roughly 4 QEC cycles (see main text). Besides the usual error sources, we also consider the contribution of the parasitic c-phase error associated with the iSWAP gates. This contributes a Pauli error of $1.5\times10^{-3}$ to the iSWAP gate, which is approximately one third of the total Pauli error. In our simulations, this error's effect was significant, accounting for $16\%$ of the total budget.

The predicted value of $\Lambda_{35}$ using the linear approximation of Eq.~\eqref{eq:1_Lambda_budget_formula} and the data in Table~\ref{Suppl:iswap-error-budget} is $\Lambda_{35}^\text{linear}=1.80$, while directly evaluating $\Lambda_{35}$ from the Pauli+ simulator gives $\Lambda_{35}^{\text{Pauli}+}=1.77$. This should be compared with the measured $\Lambda_{35}^\text{expt, harmony}=1.54$ given in the main text. As with the other codes, the simulated $\Lambda_{35}$ is larger than the measured value, indicating there are error mechanisms in the experiment not fully captured by this model. We suspect some of these mechanisms are related to the complex propagation of leakage in the code, which can be more difficult to fully capture without periodic leakage removal such as DQLR.

Finally, for reference, we show in Table~\ref{Suppl:Y1-error_budget} the $1/\Lambda$ error budget for the conventional surface code on a square lattice. 

\clearpage

\begin{table}[H]
\centering
    \begin{tabular}{|l|c|c|c|}
    \hline
      \multicolumn{1}{|c|}{Component}   & $p_{\rm expt}^{(c)}$ & $w_c$ & $\Lambda_{35}^{-1}$ contrib.\\   \hline
      CZ gates (doesn't include CZ crosstalk \& CZ leakage) & $2.25\!\times\!10^{-3}$ & 66 & 0.149 (33\%)  \\
      CZ stray coupling crosstalk & $2.5\!\times\!10^{-4}$& 131 & 0.033 (8\%)\\ 
      CZ leakage ($\ket{11}\rightarrow \ket{02}$ transition) & $2.0\!\times\!10^{-4}$ & 107 & 0.021 (5\%)\\
      Data qubit idle & $1.0\!\times\!10^{-2}$ &10 & 0.100 (23\%) \\
      Readout & $0.85\!\times\!10^{-2}$ & 7 & 0.060 (14\%) \\
      Reset & $0.15\!\times\!10^{-2}$ & 7 & 0.011 (3\%) \\
      SQ gates & $6.8\!\times\!10^{-4}$ & 85 & 0.058 (13\%)  \\
      Leakage (heating $|1\rangle$ to $|2\rangle$) & $2.6\!\times\!10^{-4}$ & 13 & 0.003 (1\%)\\
      \hline
    \end{tabular}
    \caption{{\bf Error budget for hexagonal lattice surface code.} } 
    \label{Suppl:hex-error-budget}
\end{table}

\begin{table}[H]
\centering
    \begin{tabular}{|l|c|c|c|}
    \hline
      \multicolumn{1}{|c|}{Component}   & $p_{\rm expt}^{(c)}$ & $w_c$ & $\Lambda_{35}^{-1}$ contrib.\\   \hline
      CZ gates (doesn't include CZ crosstalk \& CZ leakage) & $4.0\!\times\!10^{-3}$ & 55 & 0.218 (45\%)  \\
      CZ stray coupling crosstalk & $1.1\!\times\!10^{-4}$& 97 & 0.011 (2\%)\\ 
      CZ leakage ($\ket{11}\rightarrow \ket{02}$ transition) & $2.0\!\times\!10^{-4}$ & 140 & 0.028 (5\%)\\
      Data qubit idle & $1.0\!\times\!10^{-2}$ &6 & 0.065 (28\%) \\
      Readout & $9.4\!\times\!10^{-3}$ & 5 & 0.047 (10\%) \\
      Reset & $4.6\!\times\!10^{-3}$ & 5 & 0.023 (5\%) \\
      SQ gates & $6.6\!\times\!10^{-4}$ & 25 & 0.016 (3\%)  \\
      Leakage (heating $|1\rangle$ to $|2\rangle$) & $2.5\!\times\!10^{-4}$ & 27 & 0.007 (1\%)\\
      \hline
    \end{tabular}
    \caption{{\bf Error budget for the walking surface code.} } 
    \label{Suppl:walking-error-budget}
\end{table}

\begin{table}[H]
\centering
    \begin{tabular}{|l|c|c|c|}
    \hline
      \multicolumn{1}{|c|}{Component}   & $p_{\rm expt}^{(c)}$ & $w_c$ & $\Lambda_{35}^{-1}$ contrib.\\   \hline
      iSWAP gates (doesn't include crosstalk, leakage, or c-phase) & $2.9\!\times\!10^{-3}$ & 60 & 0.173 (31\%)  \\
      iSWAP stray coupling crosstalk & $5.6\!\times\!10^{-5}$ & 145 & 0.008 (1\%)\\ 
      iSWAP leakage & $2.1\!\times\!10^{-5}$ & 346 & 0.007 (1\%)\\
      iSWAP c-phase & $1.5\!\times\!10^{-3}$ & 58 & 0.09 (16\%)\\
      Data qubit idle & $1.3\!\times\!10^{-2}$ & 11 & 0.15 (29\%) \\
      Readout & $1.2\!\times\!10^{-2}$ & 5 & 0.062 (11\%) \\
      Reset & $1.1\!\times\!10^{-3}$ & 5 & 0.005 (1\%) \\
      SQ gates & $6.2\!\times\!10^{-4} $ & 33 & 0.02 (4\%)  \\
      Leakage (heating $|1\rangle$ to $|2\rangle$) & $2.7\!\times\!10^{-4}$ & 113 & 0.03 (5\%)\\
      \hline
    \end{tabular}
    \caption{{\bf Error budget for the iswap surface code.} } 
    \label{Suppl:iswap-error-budget}
\end{table}

\begin{table}[H]
\centering
    \begin{tabular}{|l|c|c|c|}
    \hline
      \multicolumn{1}{|c|}{Component}   & $p_{\rm expt}^{(c)}$ & $w_c$ & $(\Lambda_{3/5})^{-1}$ contrib.\\   \hline
      CZ gates (doesn't include CZ crosstalk \& CZ leakage) & $2.8\!\times\!10^{-3}$ & 65 & 0.182 (41\%)  \\
      CZ stray coupling crosstalk & $5.5\!\times\!10^{-4}$& 91 & 0.05 (11\%)\\ 
      CZ leakage ($\ket{11}\rightarrow \ket{02}$ transition) & $2.0\!\times\!10^{-4}$ & 108 & 0.022 (5\%)\\
      Data qubit idle & $0.9\!\times\!10^{-2}$ &10 & 0.09 (20\%) \\
      Readout & $0.8\!\times\!10^{-2}$ & 6 & 0.048 (11\%) \\
      Reset & $1.5\!\times\!10^{-3}$ & 6 & 0.009 (2\%) \\
      SQ gates & $6.2\!\times\!10^{-4}$ & 63 & 0.039 (9\%)  \\
      Leakage (heating $|1\rangle$ to $|2\rangle$) & $2.5\!\times\!10^{-4}$ & 18 & 0.005 (1\%)\\
      \hline
    \end{tabular}
    \caption{\bf{Error budget for the standard surface code on a square lattice.} (data from Ref.~\cite{google_quantum_ai_quantum_2025}). } 
    \label{Suppl:Y1-error_budget}
\end{table}
    \newpage

\section{Hexagonal circuit implementation of the surface code}
This section contains supplementary material for the Hexagonal implementation of the surface code.

\subsection{Circuit construction and detail}
\begin{figure}
    \centering
    \includegraphics{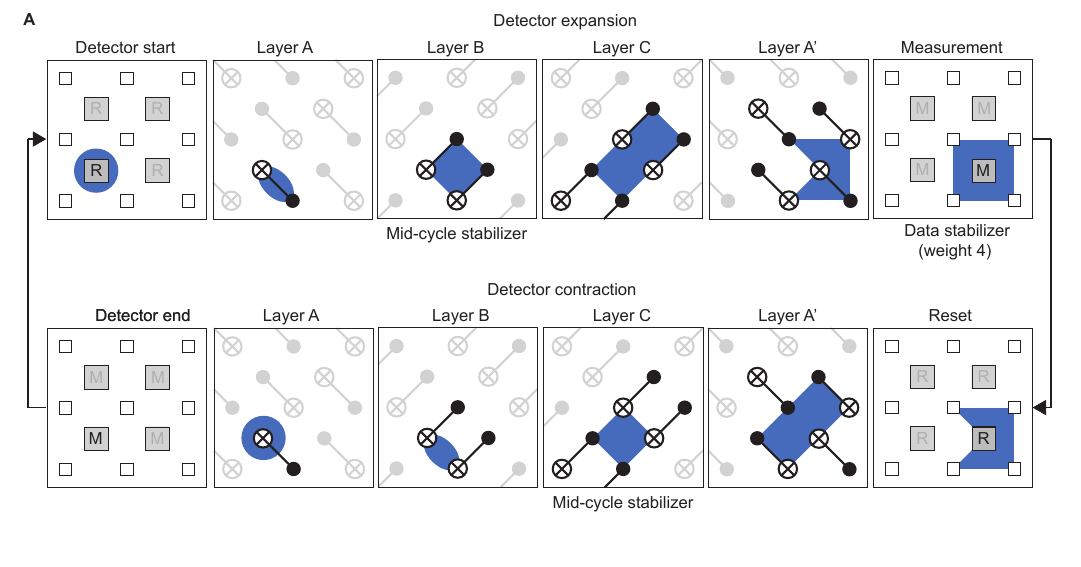}
    \caption{\textbf{Hexagonal circuit implementation of the surface code}: \textbf{a}: Single bulk detecting region for the Z-stabilizer.  Similar to the standard implementation bulk detecting region involving 2 measurements, however here the measurements are on different qubits since the detecting region moves. The sequence of entangling layers used for the hexagonal implementation reuses the same connectivity of the first and fourth layer each cycle with the direction of the CNOT inversed. This inversing is done in hardware with single qubit gates. Each bulk detecting region touches 22 CNOT gates, similar to the standard surface code.}
    \label{fig:supp_hex_circuit}
\end{figure}

The circuit construction for the hexagonal circuit implementation of the surface code follows the recipe given in \cite{mcewen_relaxing_2023}. 
We reproduce here its construction while giving insight into the construction's impact on the error budget. 
We present the hexagonal implementation circuit in ~\fig{supp_hex_circuit} in terms of CNOT gates for clarity of presentation. 
In our experiments, the CZ gate is used instead, and single-qubit rotations (Hadamards) are inserted into the circuit to change basis.

In ~\fig{supp_hex_circuit}, we show the evolution of a single Z-detecting region through the full circuit. 
The detection region first goes through a expansion cycle then a contraction cycle. For this implementation, the contraction cycle is the time inverse of the expansion cycle. 
Note that the measurements in the detecting region are not on the same qubits due to the detecting region's movement during the circuit. Finally, the last layer before measurement is the same entangling layer as the first entangling layer of the cycle with the direction if the CNOT inversed. This property is what allows the use only hexagonal connectivity.
This representation highlights that each bulk detecting region touches 22 differents CNOTs, similar to the standard surface code, however, the detecting region touches less qubits (only 6 compared to 9 for the standard surface code), and less CZ gates when the CNOT is decomposed into Hadamard and CZ.

\begin{figure}[h]
    \centering
    \includegraphics[width=\textwidth]{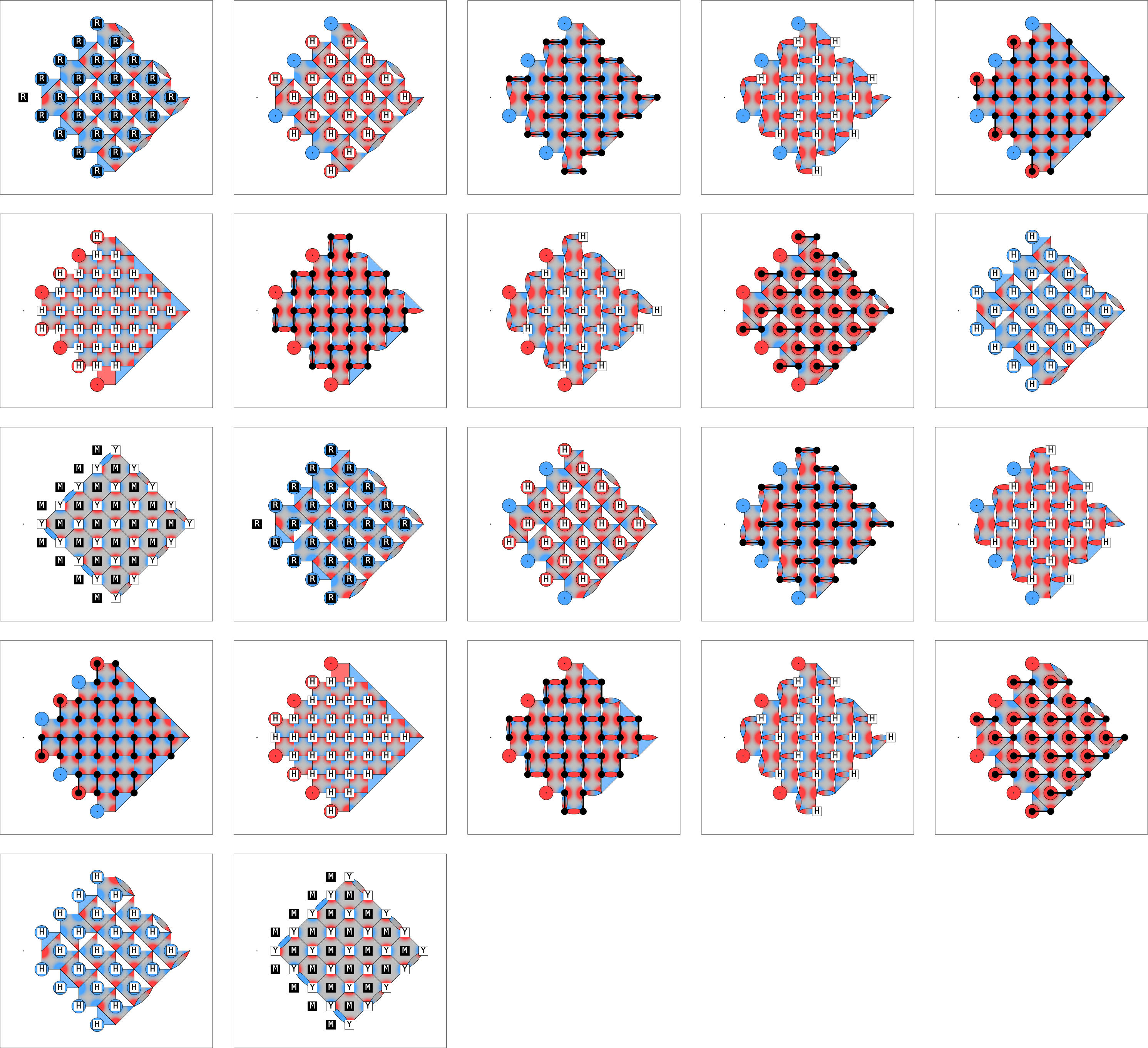}
    \caption{
    \textbf{Hexagonal circuit detector slices}: 
    Each gate layer in two consecutive bulk rounds in the circuit are shown, with the detecting region slices immediately after that gate layer shown in the background.
    Each shape indicates a single detecting region slice, with non-identity terms on qubits at the vertices of that shape.
    The color at each vertex indicates the Pauli term, where (Red, Blue, Green) indicates (X, Y, Z) respectively.
    The reset gates (R) represent the Pauli action of the multi-level reset and DQLR. 
    The additional gates during measurement (Y) represent the Pauli action of the dynamical decoupling sequence.
    The full circuit can be explored interactively in Crumble: \href{\hexcrumble}{Click Here.}
    }
    \label{fig:supp_hex_full_slice}
\end{figure}

In ~\fig{supp_hex_full_slice}, we show the full detector slice of the circuit implemented on the hardware with all the compiled Hadamards. Note that at the reset slice, stabilizer at the north-west boundary span two measure qubits. This is necessary for the boundary of the surface code to be valid. This makes the edge of this code different from the standard implementation and also changes the noise sensitivity. 

\subsection{Component Benchmarks}
\label{subsec:supp_hex_benchmarks}
\begin{figure}[h]
    \centering
    \includegraphics{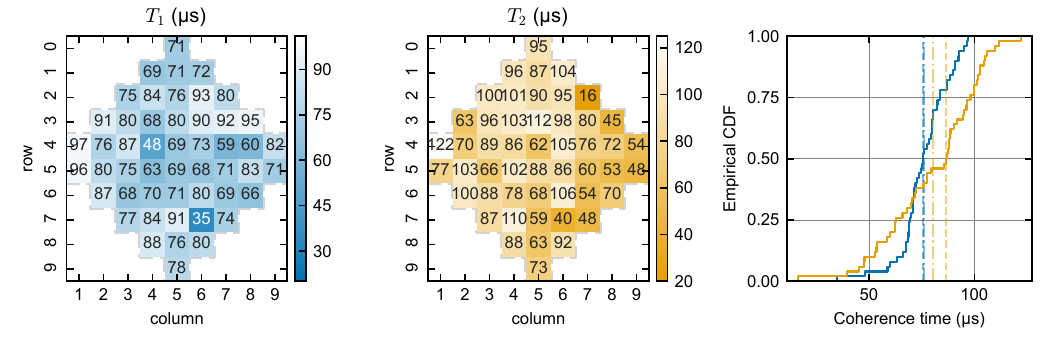}
    \caption{\textbf{Hexagonal grid, coherence}. \textbf{a}: T1 and \textbf{a}: T2 of the hexagonal grid measure with interleaved population decay experiment and echo experiments. We measure the following statistical quantities for $T_1$: mean: $76$ $\mu$s, median: $76$ $\mu$s, std:  $12$ $\mu$s and for for $T_2$: mean: $80$ $\mu$s, median: $87$ $\mu$s, std:  $22$ $\mu$s     }
    \label{fig:supp_hex_coherences}
\end{figure}

\begin{figure}[h]
    \centering
    \includegraphics{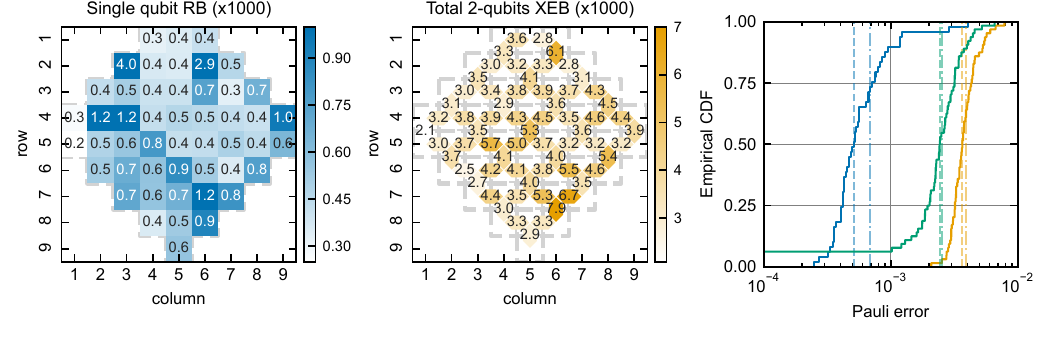}
    \caption{\textbf{Hexagonal grid, Randomized Benchmarking}. \textbf{a}: Single qubit Pauli error measured through randomized benchmarking. \textbf{b}: Two qubit XEB Pauli error measured with XEB. Here we report the total error that include the single qubit error. \textbf{c}: Empirical cumulative histogram of the Pauli error rates across the chip. The green represents the \textit{inferred} error of the CZ where the single qubit gate error (blue) has been subtracted from the total XEB error (yellow). }
    \label{fig:supp_hex_rb}
\end{figure}

\begin{figure}[h]
    \centering
    \includegraphics{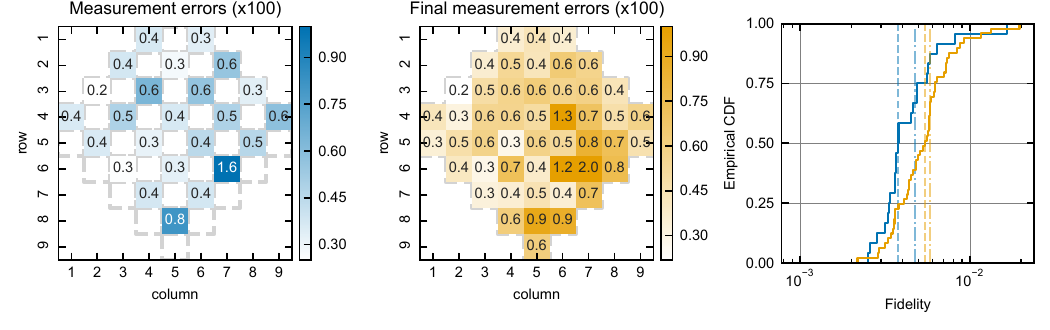}
    \caption{\textbf{Hexagonal grid, Readout Benchmarking}. \textbf{a}: Average readout error during a mid-cycle measurement round using randomized state readout benchmarking. \textbf{b}: Same experiment for the final round of the memory experiment where all qubits are measured. \textbf{c}: Empirical cumulative histogram of the readout errors for both mid-circuit and termination measurements.}
    \label{fig:supp_hex_readout}
\end{figure}

\begin{figure}[h]
    \centering
    \includegraphics{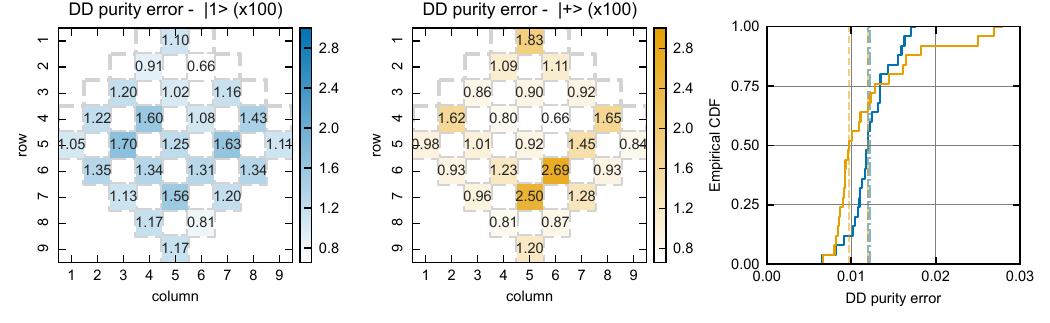}
    \caption{\textbf{Hexagonal implementation, Dynamical Decoupling}. \textbf{a}: Purity decay (or population decay) starting from a $\ket{1}$ state during the measurement and reset operation in a memory experiment. \textbf{b}: Purity decay (or population decay) starting from a $\ket{+}$ state during the measurement and reset operation in a memory experiment. \textbf{c}: Empirical cumulative histogram of the dynamical decoupling errors during the readout moment.}
    \label{fig:supp_hex_dd}
\end{figure}

\begin{figure}[h]
    \centering
    \includegraphics{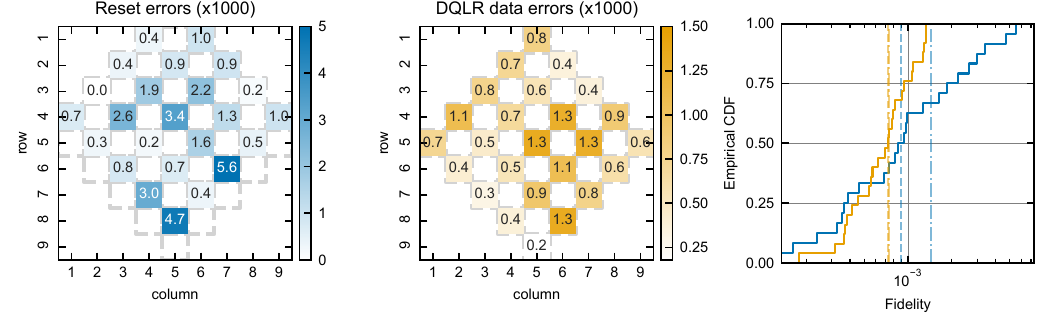}
    \caption{\textbf{Hexagonal implementation, Reset and Leakage Removal}. \textbf{a}: Error (residual one state) after a reset on a measurement qubit. \textbf{b}: Excess depolarizing error on the data qubit during the Data-Qubit Leakage Removal (DQLR) step. \textbf{c}: Empirical cumulative histogram of the Reset and DQLR error rates. }
    \label{fig:supp_hex_reset}
\end{figure}

In this subsection, we report all the component benchmark measured for the hexagonal implementation. ~\fig{supp_hex_coherences} shows the coherence times of the grid used for implementing the hexagonal implementation. ~\fig{supp_hex_rb} shows the raw gate benchmarks with single qubit RB and two-qubits XEB. ~\fig{supp_hex_readout} shows the readout benchmark for the mid-circuit and terminal measurements. ~\fig{supp_hex_dd} shows the idling error during the measurement cycles (or error during the Dynamical Decoupling sequence). ~\fig{supp_hex_reset} shows reset errors. All the component benchmarks are summarized in table~\ref{tab:component_benchmarks}.

\begin{table}[hbt!]
    \centering
    \begin{tabular}{|c|c|c|c|c|c|c|c|c|c|c|c|}
        \hline
        & count & mean & median & std & q1 & q3 & min & max & IQR & lower outliers & upper outliers \\
        \hline \hline
        T1 (\SI{}{\micro\second}) & 50 & 76.15 & 75.63 & 11.82 & 69.40 & 83.28 & 34.69 & 96.87 & 13.88  & 2 & 0  \\
        T2 (\SI{}{\micro\second}) & 50 & 80.43 & 86.67 & 21.85 & 63.60 & 97.71 & 16.30 & 122.05 & 34.10 & 0 & 0  \\
        Single Qubit RB ($\times 10^3$) & 49 & 0.69 & 0.51 & 0.63 & 0.42 & 0.72 & 0.25 & 4.05 & 0.30 & 0 & 5  \\
        Two qubits XEB (total $\times 10^3$) & 64 & 3.88 & 3.66 & 1.01 & 3.21 & 4.30 & 2.08 & 7.87 & 1.09 & 0 & 3  \\
        Two qubits XEB (inferred $\times 10^3$) & 64 & 2.55 & 2.43 & 1.16 & 2.00 & 3.18 & 0.00 & 6.59 & 1.17 & 4 & 2  \\
        Mid-cycle readout ($\times 10^2$) & 24 & 0.48 & 0.38 & 0.28 & 0.34 & 0.51 & 0.25 & 1.65 & 0.17 & 0 & 2  \\
        Mid-cycle reset ($\times 10^3$) & 24 & 1.44 & 0.89 & 1.43 & 0.38 & 1.95 & 0.00 & 5.56 & 1.57 & 0 & 2 \\
        Final measurement ($\times 10^2$) & 49 & 0.58 & 0.54 & 0.29 & 0.41 & 0.63 & 0.22 & 1.96 & 0.22 & 0 & 3  \\
        Dynamical Decoding (T1) purity ($\times 10^2$) & 25 & 1.22 & 1.20 & 0.24 & 1.10 & 1.34 & 0.66 & 1.70 & 0.25 & 1 & 0  \\
        Dynamical Decoding (T2) purity ($\times 10^2$) & 25 & 1.20 & 0.98 & 0.50 & 0.90 & 1.28 & 0.66 & 2.69 & 0.38 & 0 & 2  \\
        DQLR excess Pauli error ($\times 10^3$) & 25 & 0.74 & 0.73 & 0.34 & 0.45 & 0.95 & 0.18 & 1.33 & 0.49 & 0& 0  \\
        \hline
    \end{tabular}
    \caption{\textbf{Hexagonal implementation benchmarking.} Statistical description of the component benchmarks with Pauli error rates. Outlier are counted as point outside the 1.5 IQR range.}
    \label{tab:component_benchmarks}
\end{table}


\clearpage
\subsection{Detection analysis}
In this section, we perform a detailed analysis of the detection events experimentally measured during the experiment on the hexagonal implementation. We mainly focus on the distance 5 hexagonal circuit with 54 cycles to confirm that our Pauli models used for the detection probability budget properly predict the experimental data. 

\begin{figure}[h]
    \centering
    \includegraphics{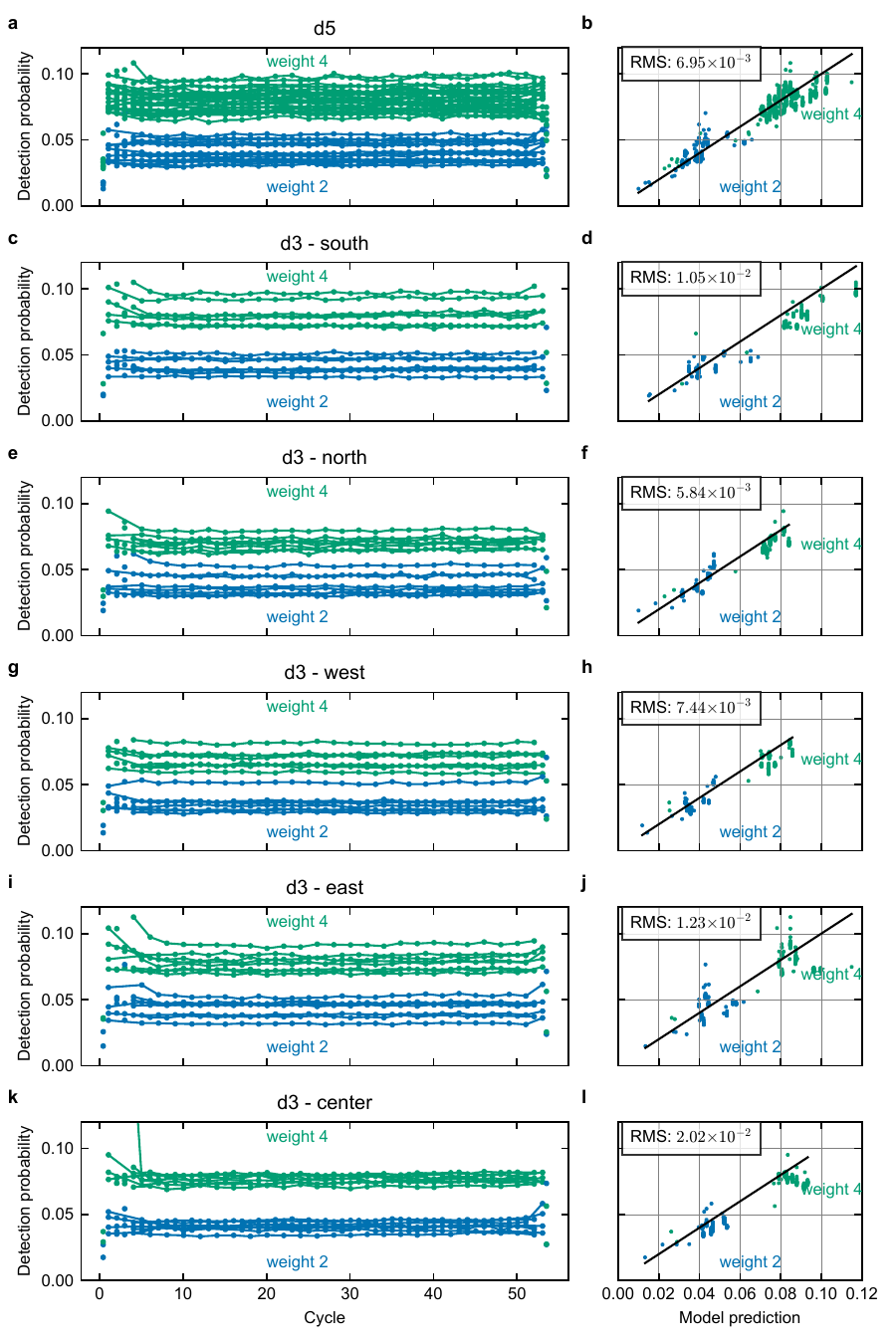}
    \caption{\textbf{Hexagonal implementation, Detection Fraction analysis}: Left panels show the average detection fraction of all experiment in the Z basis. The right panels shows the correlation of the experimental detection probabilities with the models predictions.}
    \label{fig:supp_hex_det_frac}
\end{figure}

Our first step is to look at the detection fraction (of first statistical moment) of the detection events. In Fig~\ref{fig:supp_hex_det_frac}\textbf{a}, we show the average of each detectors over 20 different initial bitstrings and 10k repetitions per experiment. We see a clear separation between the boundary (weight 2) and the bulk (weight 4) detectors. 
In Fig~\ref{fig:supp_hex_det_frac}\textbf{b}, we show the correlation of these detection fractions with simulations. We also present the Root Mean Square (RMS) difference of the measured detection probabilities with the Pauli error model constructed from the benchmarks of the previous section \ref{subsec:supp_hex_benchmarks}. The explicit formula is:
\begin{equation}
    RMS = \sqrt{\frac{1}{N_{\text{detector}}} \sum_{\text{shots}} (d_{meas} - d_{sim})^2}
\end{equation}
where the value of the detectors is average over the shots. This quantity allow for assessing the predictibity or our Pauli model.

We see that even though the benchmarks have been taken in the context of the distance-5 memory, the Pauli error model still captures the detection probabilities for the embedded distance-3 codes.

\begin{figure}[h]
    \centering
    \includegraphics{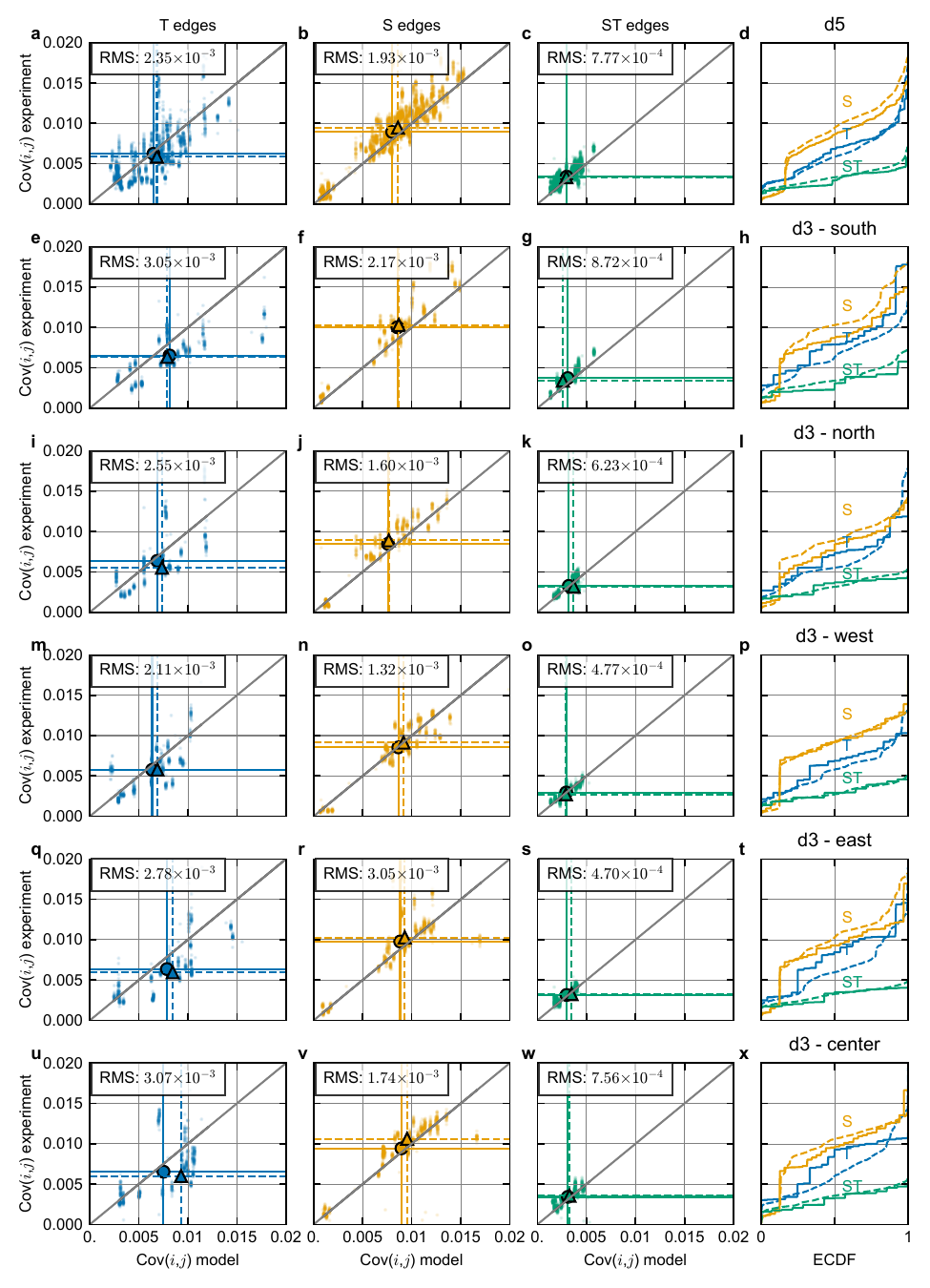}
    \caption{\textbf{Hexagonal implementation, Detector covariance analysis -  regular edges}: \textbf{a column}: Space-like (S) edges \textbf{b column}: Timelike (T) edges \textbf{c column}: Spacetimelike (ST) edges \textbf{d column}: Cumulative histogram of the edges}
    \label{fig:supp_hex_covariance}
\end{figure}

In Fig~\ref{fig:supp_hex_covariance}, we show the correlation of the covariance term for pairs of detectors corresponding to edges in the decoding graph. Note that we only show the 2-point covariance. We use the usual categorization where T-edges refer to detectors that share the same measurements but are shifted in time. S edges denote detectors that share data qubits during the measurement moment i.e. detectors that span the same cycles but touche on a few qubits. Finally ST edges correspond to pair of detectors that are shifted in time but still touch through CZ gates layers. We also present the RMS of the measured covariance with the Pauli error model constructed from the benchmarks of the previous section \ref{subsec:supp_hex_benchmarks} and using the formula from \ref{sec:supp_stats}. We also see a very good agreement between the the predicted values from the Pauli model and the measured values for these edges.

\clearpage
    \section{Walking circuit implementation of the surface code}
This section contains supplementary material for the walking implementation of the surface code.

\subsection{Circuit construction and detail}
\begin{figure}[hb]
    \centering
    \includegraphics[width=\textwidth]{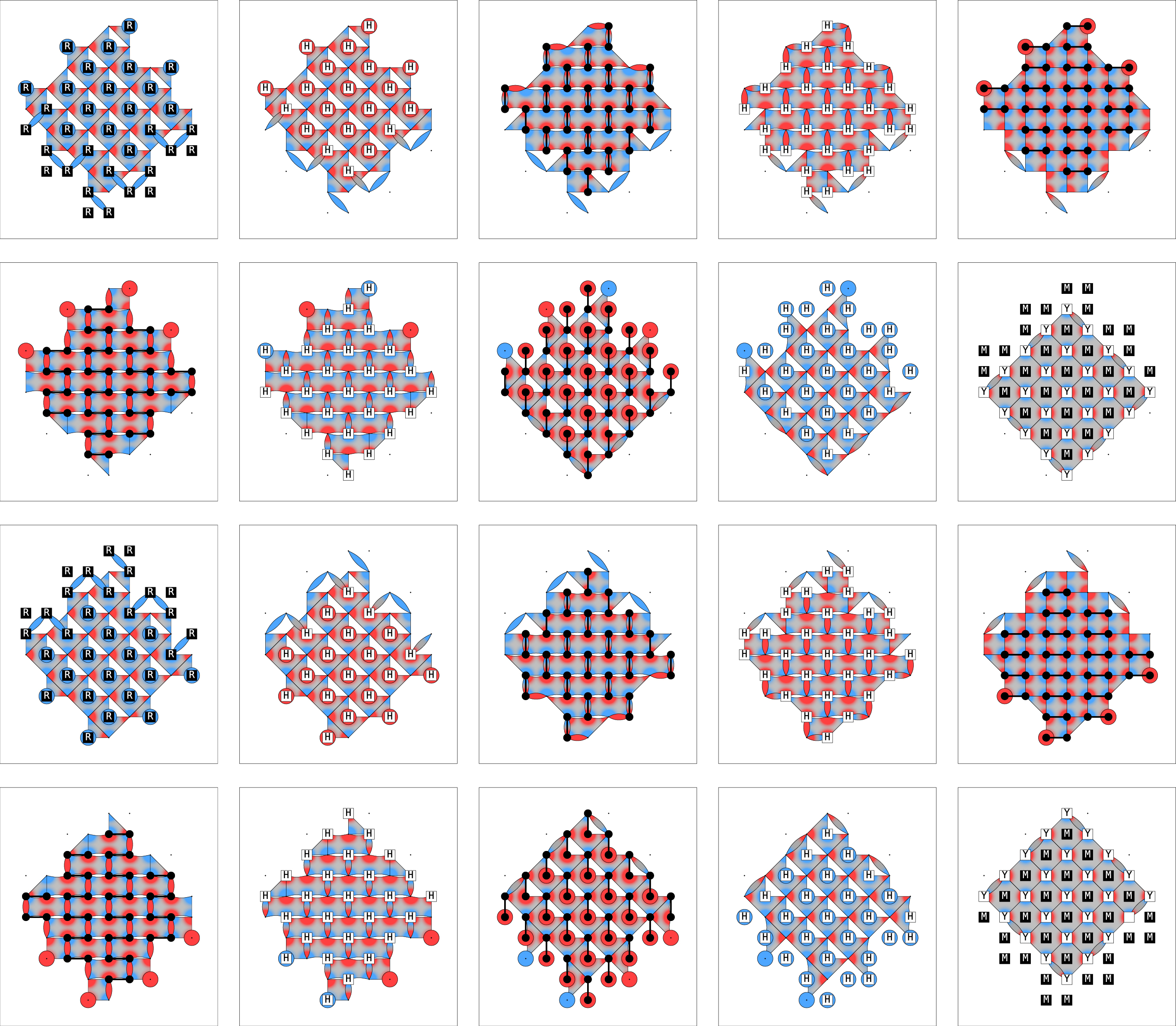}
    \caption{
    \textbf{Walking circuit detector slices}: 
    Each gate layer in two consecutive bulk rounds in the circuit are shown, with the detecting region slices immediately after that gate layer shown in the background.
    Each shape indicates a single detecting region slice, with non-identity terms on qubits at the vertices of that shape.
    The color at each vertex indicates the Pauli term, where (Red, Blue, Green) indicates (X, Y, Z) respectively.
    The additional gates during measurement (Y) represent the Pauli action of the dynamical decoupling sequence.
    The full circuit can be explored interactively in Crumble: \href{\walkcrumble}{Crumble link to interactive circuit for the Walking implementation.}
    }
    \label{fig:supp_walk_full_slice}
\end{figure}

The circuit construction for the walking circuit implementation of the surface code follows the proposal from \cite{mcewen_relaxing_2023}. In ~\fig{supp_walk_full_slice} we show the full detector slices for the 2-cycles of the walking implementation we have used in the main text, including the Hadamards. The circuit is obtained by a space-inversion between the first cycle and the second, hence every other cycle is spatially symmetric to the previous one. This allows the inclusion of all qubits (including all boundaries one) in the reset layer to eliminate all leakage error. Note that using the time-reversal circuit for the contraction of the detector also yield a functioning circuit. However, this circuit does not implement a reset on all the qubits (on the boundaries) and can allow leakage accumulation on qubits that are not reset every cycle, hence our choice for space symmetry for this specific implementation.
Note that one could walk a qubit to a fully different location using this circuit as a primitive for the operation of moving a logical qubit.

\subsection{Component benchmarks}
\label{subsec:supp_walking_benchmarks}
\begin{figure}[h]
    \centering
    \includegraphics{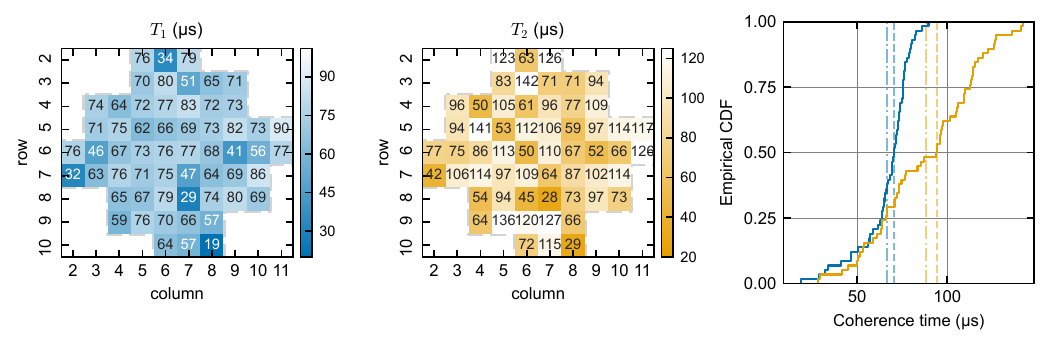}
    \caption{\textbf{Walking implementation, coherence}. \textbf{a}: T1 and \textbf{b}: T2 of the walking grid measure with interleaved population decay experiment and echo experiment. We measure the following statistical quantities for $T_1$: mean: $67$ $\mu$s, median: $70$ $\mu$s, std:  $14$ $\mu$s and for for $T_2$: mean: $88$ $\mu$s, median: $94$ $\mu$s, std:  $28$ $\mu$s     }
    \label{fig:supp_walking_coherences}
\end{figure}

\begin{figure}[h]
    \centering
    \includegraphics{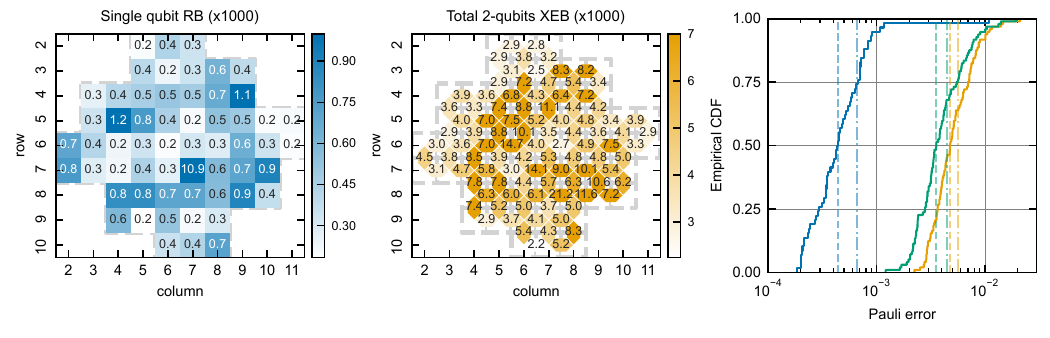}
    \caption{\textbf{Walking implementation, Randomized Benchmarking}. \textbf{a}: Single qubit Pauli error measured through randomized benchmarking. \textbf{b}: Two qubit XEB Pauli error measured with XEB. Here we report the total error that include the single qubit error. \textbf{c}: Empirical cumulative histogram of the Pauli error rates across the chip. The green represents the \textit{inferred} error of the CZ where the single qubit gate error (blue) has been subtracted from the total XEB error (yellow). }
    \label{fig:supp_walking_rb}
\end{figure}

\begin{figure}[h]
    \centering
    \includegraphics{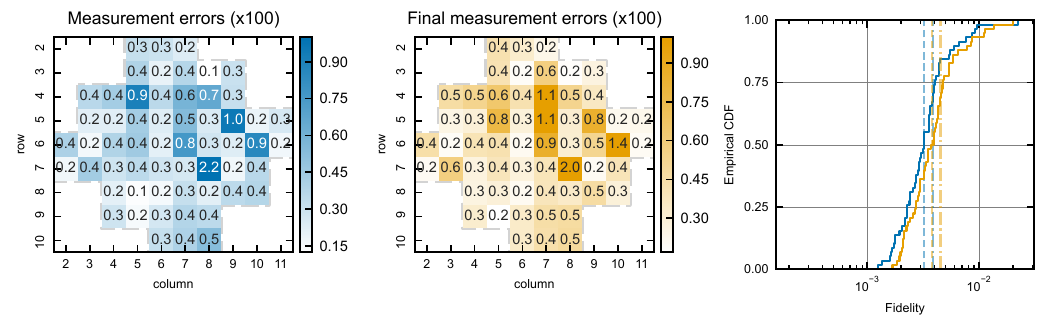}
    \caption{\textbf{Walking implementation, Readout Benchmarking}. \textbf{a}: Average readout error during a mid-cycle measurement round using randomized state readout benchmarking. \textbf{b}: Same experiment for the final round of the memory experiment where all qubits are measured. \textbf{c}: Empirical cumulative histogram of the readout errors for both mid-circuit and termination measurements. }
    \label{fig:supp_walking_readout}
\end{figure}

\begin{figure}[h]
    \centering
    \includegraphics{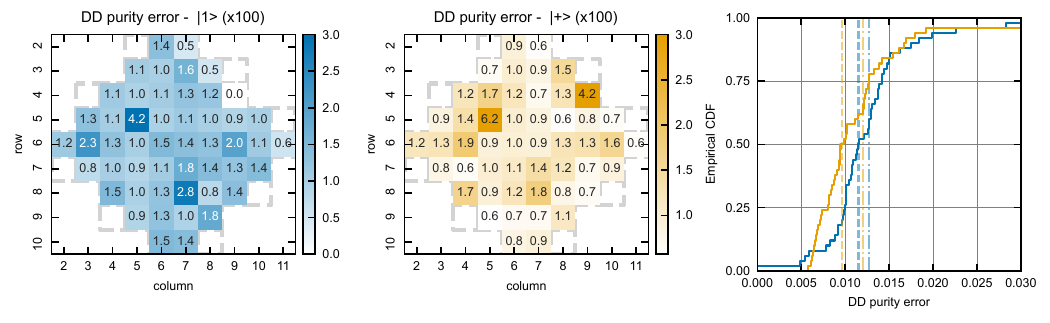}
    \caption{\textbf{Walking implementation, Dynamical Decoupling}.  \textbf{a}: Purity decay (or population decay) starting from a $\ket{1}$ state during the measurement and reset operation in a memory experiment. \textbf{b}: Purity decay (or population decay) starting from a $\ket{+}$ state during the measurement and reset operation in a memory experiment. \textbf{c}: Empirical cumulative histogram of the dynamical decoupling errors during the readout moment.}
    \label{fig:supp_walking_dd}
\end{figure}

\begin{figure}[h]
    \centering
    \includegraphics{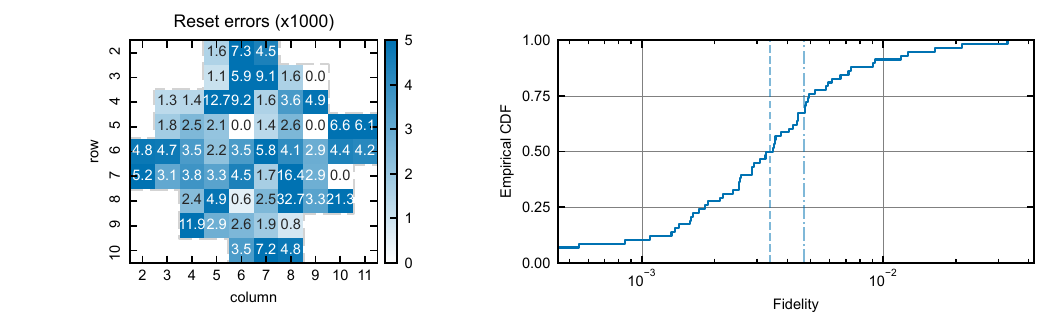}
    \caption{\textbf{Walking implementation Reset}. \textbf{a}: Error (residual one state) after a reset on a measurement qubit.  \textbf{b}: Empirical cumulative histogram of the Reset  error rates. }
    \label{fig:supp_walking_reset}
\end{figure}

In this subsection, we report all the component benchmark measured for the walking implementation. ~\fig{supp_walking_coherences} shows the coherence times of the grid used. ~\fig{supp_walking_rb} shows the raw gate benchmarks with single qubit RB and two-qubits XEB. ~\fig{supp_walking_readout} shows the readout benchmark for the mid-circuit and terminal measurements. ~\fig{supp_walking_dd} shows the idling error during the measurement cycles (or error during the Dynamical Decoupling sequence). ~\fig{supp_walking_reset} shows reset errors. All the component benchmarks are summarized in table~\ref{tab:component_benchmarks_walking}.

The walking implementation was done on the 105-qubit Williow device used for the distance-7 demonstration in \cite{google_quantum_ai_quantum_2025}. During the calibration process, the qubit at the position (7, 8) presented a low-fidelity multi-level reset operation. This can be seen in the reset benchmark \fig{supp_walking_reset}. Since all qubits need to be reset at some point in this implementation, this defective reset has large impact on the code performance. In the distance-7 experiment of \cite{google_quantum_ai_quantum_2025}, this qubit was used as a data qubit and hence it poor reset fidelity didn't impact its performance. In our walking implementation here, we replaced the multi-level reset on qubit (7,8) only with a single-level reset, which had higher fidelity for reset of the $\ket{1}$ state. The fact that the walking implementation still achieved high performance with this change is an indication of the leakage benefits to walking.

\begin{table}[h]
  \centering
  \begin{tabular}{|c|c|c|c|c|c|c|c|c|c|c|c|}
  \hline
   & count & mean & median & std & q1 & q3 & min & max & IQR & low outliers & high outliers \\
  \hline
  T1 (\SI{}{\micro\second}) & 58 & 66.74 & 70.43 & 14.06 & 63.64 & 75.55 & 19.08 & 89.56 & 11.91 & 5 & 0  \\
  T2 (\SI{}{\micro\second}) & 58 & 88.11 & 94.04 & 28.35 & 65.91 & 111.72 & 28.27 & 141.91 & 45.81 & 0 & 0  \\
  Single Qubit RB ($\times 10^3$) & 58 & 0.66 & 0.44 & 1.38 & 0.30 & 0.66 & 0.18 & 10.94 & 0.37 & 0 & 1  \\
  Two qubits XEB (total $\times 10^3$) & 97 & 5.63 & 4.77 & 2.94 & 3.66 & 7.02 & 2.24 & 21.19 & 3.35 & 0 & 3  \\
  Two qubits XEB (inferred $\times 10^3$) & 97 & 4.49 & 3.56 & 2.83 & 2.81 & 5.52 & 1.21 & 19.38 & 2.71 & 0 & 5  \\
  Mid-cycle readout ($\times 10^2$) & 58 & 0.39 & 0.32 & 0.30 & 0.23 & 0.40 & 0.12 & 2.22 & 0.17 & 0 & 6  \\
  Mid-cycle reset ($\times 10^3$) & 58& 4.70 & 3.39 & 5.34 & 1.84 & 4.91 & 0.00 & 32.73 & 3.08 & 0 & 5 \\
  Final measurement ($\times 10^2$)& 58 & 0.45 & 0.38 & 0.31 & 0.27 & 0.47 & 0.17 & 2.00 & 0.20 & 0 & 6  \\
  Dynamical Decoding (T1) purity ($\times 10^2$) & 50 & 1.27 & 1.15 & 0.60 & 1.00 & 1.41 & 0.00 & 4.16 & 0.41 & 1 & 3  \\
  Dynamical Decoding (T2) purity ($\times 10^2$) & 50 & 1.21 & 0.97 & 0.90 & 0.81 & 1.27 & 0.57 & 6.20 & 0.46 & 0 & 2  \\
  \hline
  \end{tabular}
  \caption{\textbf{Walking implementation benchmarking.} Statistical description of the component benchmarks with Pauli error rates. Outlier are counted as point outside the 1.5 IQR range.}
 \label{tab:component_benchmarks_walking}
\end{table}

\clearpage
\subsection{Detection analysis}
\begin{figure}[h]
    \centering
    \includegraphics{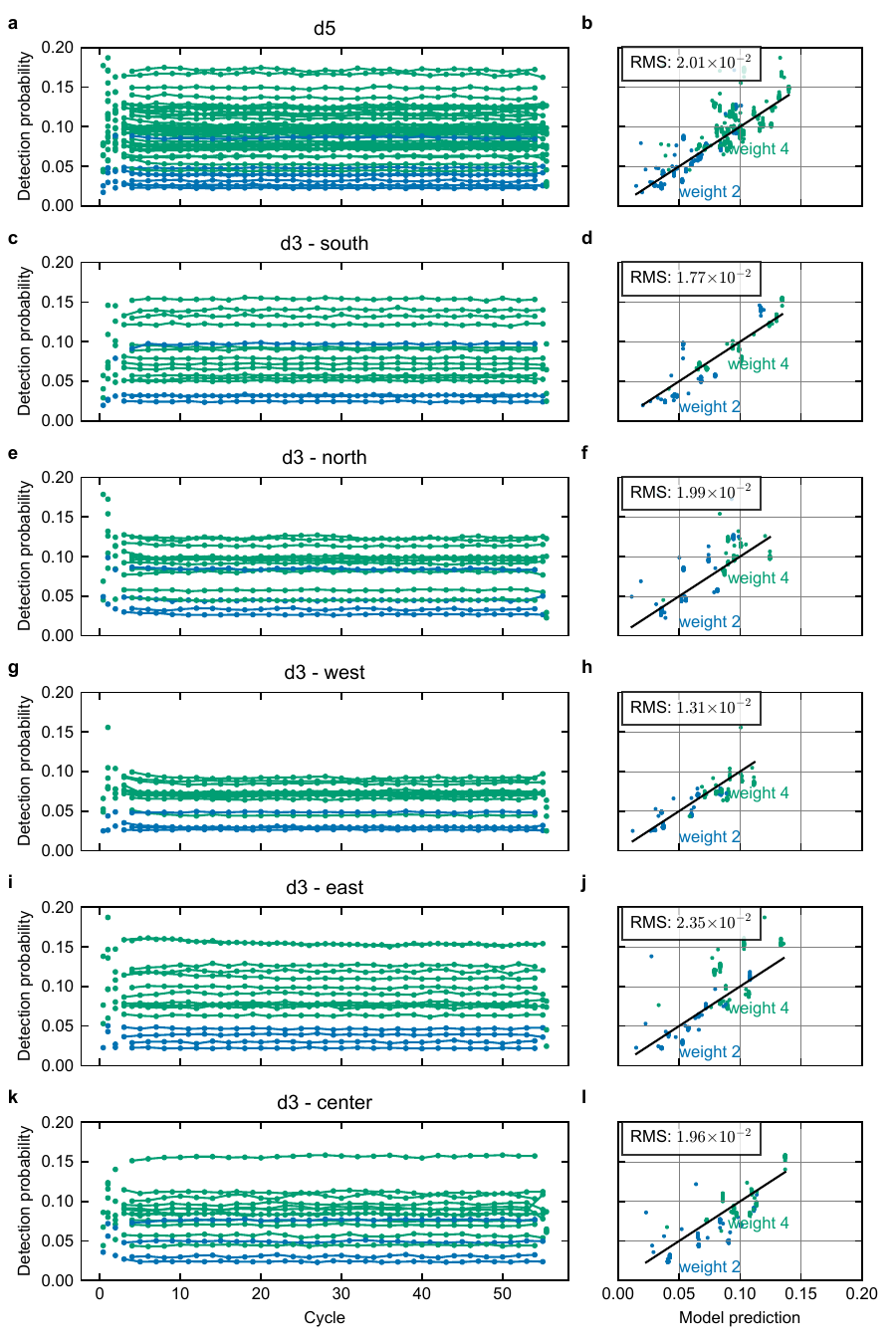}
    \caption{\textbf{Walking implementation, Detection Fraction analysis}: Left panels show the average detection fraction of all experiment in the Z basis. The right panels shows the correlation of the experimental detection probabilities with the models predictions.}
    \label{fig:supp_walking_det_frac}
\end{figure}

In this section, we present a detailed analysis of the detection events experimentally collected during the experiment on the walking implementation. We mainly focus on the distance 5 and distance 3 walking circuits with 54 cycles to confirm that our Pauli  models properly predict the experimental data. Our first step is to look at the detection fraction (of first statistical moment) of the detection events. In ~\fig{supp_walking_det_frac}\textbf{a}, we show the average of each detectors over 20 different initial bitstrings and 10k repetitions per experiments. We see a clear separation between the boundary (weight 2) and the bulk (weight 4) detectors. 
In \fig{supp_walking_det_frac}\textbf{b}, we show the correlation of these detection fraction from Pauli simulations used to construct the detector probability error budgets. We also present the RMS of the measured detection probabilities with the Pauli error model constructed from the benchmarks of the previous section \ref{subsec:supp_walking_benchmarks}. We see that even-though the benchmarks have been taken in the context of the distance-5 memory, the Pauli error model still capture the detection probabilities for the embedded distance-3 codes. However, the Pauli models for the walking code have a weaker agreement with measured data than for the hexagonal or iSWAP implementations. 

\begin{figure}[h]
    \centering
    \includegraphics{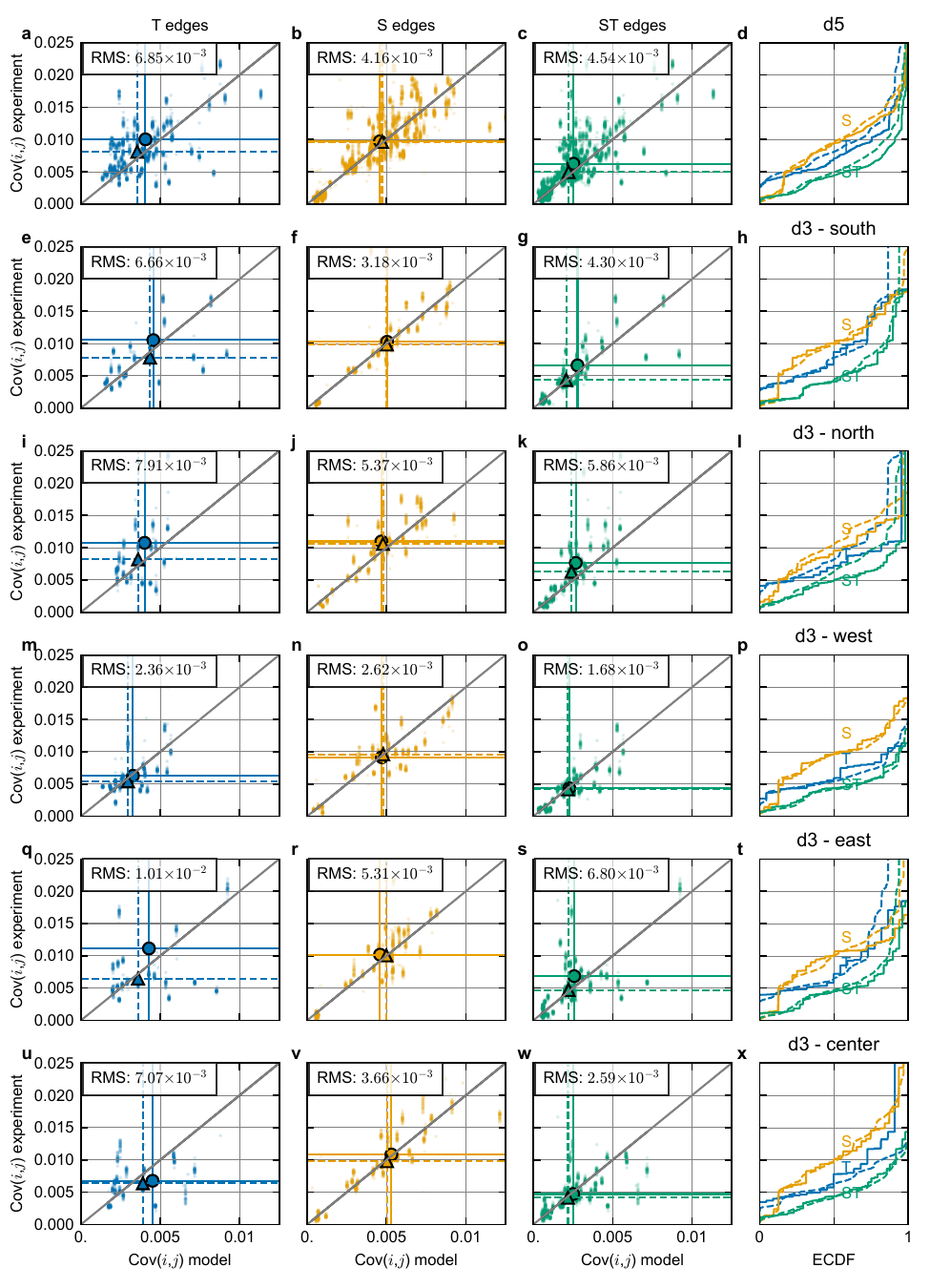}
    \caption{\textbf{Walking implementation, Detector covariance analysis -  regular edges}. \textbf{a column}: Space-like edges \textbf{b column}: Time-like edges \textbf{c column}: spacetime-like edges \textbf{d column}: Cumulative histogram of the edges}
    \label{fig:supp_walking_covariance}
\end{figure}

    \section{iSWAP circuit implementation of the surface code}
This section contains supplementary material for the iSWAP implementation of the surface code.

\subsection{Circuit construction and detail}
\begin{figure}[hb]
    \centering
    \includegraphics[width=\textwidth]{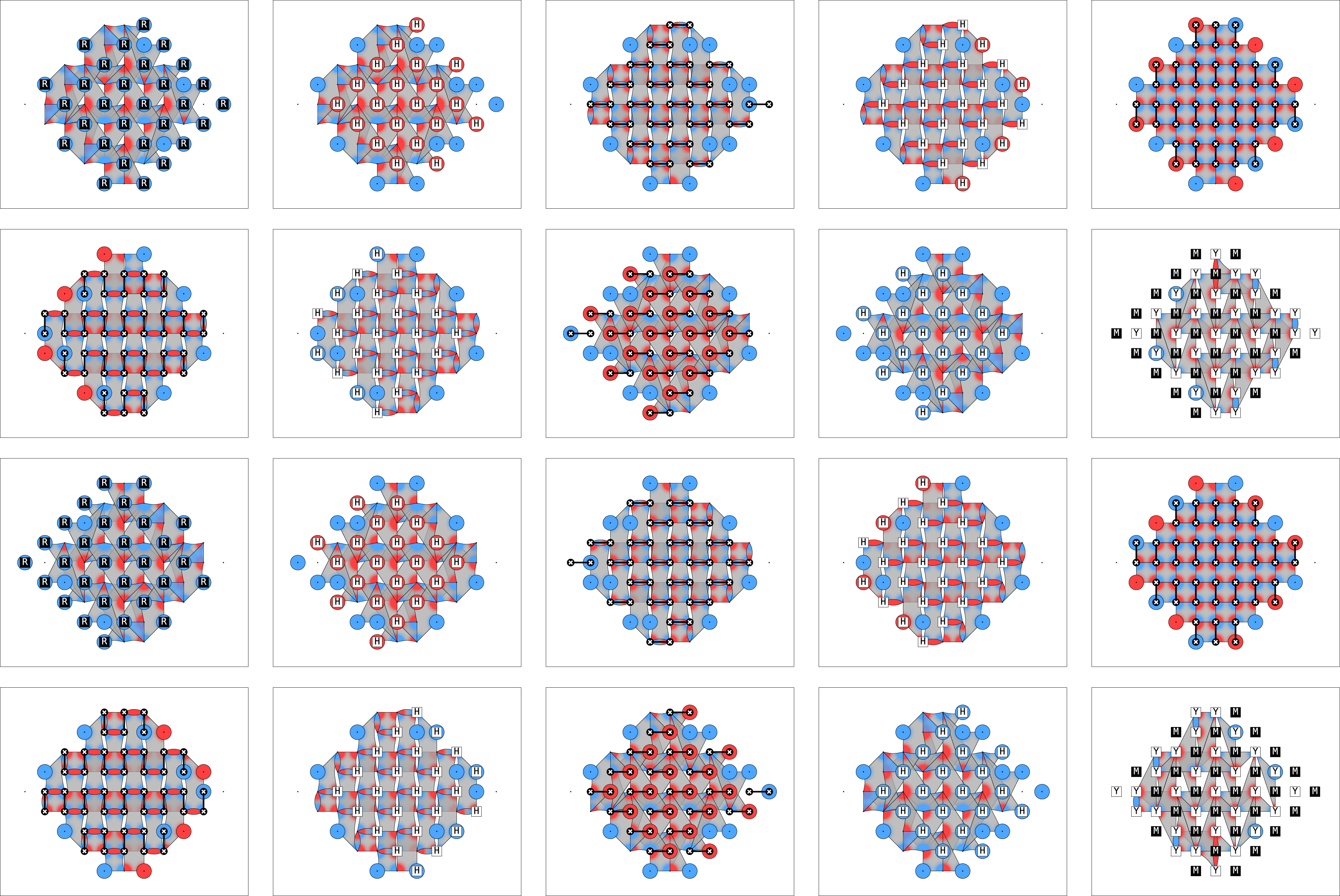}
    \caption{
    \textbf{iSWAP circuit detector slices}: 
    Each gate layer in two consecutive bulk rounds in the circuit are shown, with the detecting region slices immediately after that gate layer shown in the background.
    Each shape indicates a single detecting region slice, with non-identity terms on qubits at the vertices of that shape.
    The color at each vertex indicates the Pauli term, where (Red, Blue, Green) indicates (X, Y, Z) respectively.
    The two-qubit gates indicated are CZSWAP gates, equivalent to CZ followed by SWAP, and equivalent to iSWAP up to single qubit Clifford gates.
    The additional gates during measurement (Y) represent the Pauli action of the dynamical decoupling sequence.
    The full circuit can be explored interactively in Crumble: \href{\iswapcrumble}{Crumble for the iSWAP implementation.}
    }
    \label{fig:supp_iswap_full_slice}
\end{figure}

In ~\fig{supp_iswap_full_slice} we show the full detector slices for 2-cycle iSWAP implementation we have used in the main text, including the Hadamards. The circuit is obtained by first constructing the four layers of iSWAP similar to a CNOT circuit. While doing so, the detectors are more spread out than with CNOT/CZ gates. However, using the time-reversal in the second cycle allows their refocusing. The boundary of the code is significantly harder to build and have necessitated some trial and error to yield a functioning code. We give a Crumble link for interactive exploration of the circuit here: \href{\iswapcrumble}{Crumble for the iSWAP implementation.}.

\subsection{Experimental implementation details}

\textbf{Balanced-Sycamore gate:} We implement the iSWAP gate using a $01$-$10$ interaction between two flux-tunable transmons. This interaction yields an unwanted c-phase component due to the interaction of higher energy states. This c-phase introduces a new error channel for QEC codes that require strict Cliffords. To minimize the impact of this c-phase, we \textit{balance} the c-phase using single qubit virtual Z rotations similar to \cite{mi_information_2021}. Specifically, the gate unitary we implement on hardware (using $\gamma = -\phi/2$) is:

\begin{equation}
    U = \begin{bmatrix} 
        1 & 0 & 0 & 0 \\
        0 & 0 & ie^{i\gamma} & 0 \\
        0 & ie^{i\gamma} & 0 & 0 \\
        0 & 0 & 0 & e^{i(2\gamma + \phi)}
    \end{bmatrix}
    = \begin{bmatrix} 
     1 & 0 & 0 & 0 \\
     0 & 0 & i e^{-\frac{i \phi }{2}} & 0 \\
     0 & i e^{-\frac{i \phi }{2}} & 0 & 0 \\
     0 & 0 & 0 & 1 \\
    \end{bmatrix}
\end{equation}

We call this gate the balanced-Sycamore gate since we are balancing the single qubit Z phase. Using this gate instead of the Sycamore gate directly allow the reduction of error associated with the cphase by a factor 3 from $\varepsilon =\frac{3}{16} \phi^2$ to $\varepsilon =\frac{1}{16} \phi^2$. For our simulation, since the c-phase is a coherent non-Clifford unitary effect, we approximate its effect to a Pauli-stochastic channel. By comparing incoherent simulations with coherent simulations of the distance-3 iSWAP surface code, we have confirmed that this Pauli-twirling of this coherent error is a valid approximation in the code. In our Pauli simulations, we then model the coherent error using the channel: 
\begin{equation*}
\begin{aligned}
    &\rho \rightarrow \cos^2\left(\frac{\phi}{4}\right) \rho + \sin^2\left(\frac{\phi}{4}\right) ZZ\rho ZZ.
\end{aligned}
\end{equation*}

\subsection{Component Benchmarks}
\label{subsec:supp_iSWAP_benchmarks}
\begin{figure}[h]
    \centering
    \includegraphics{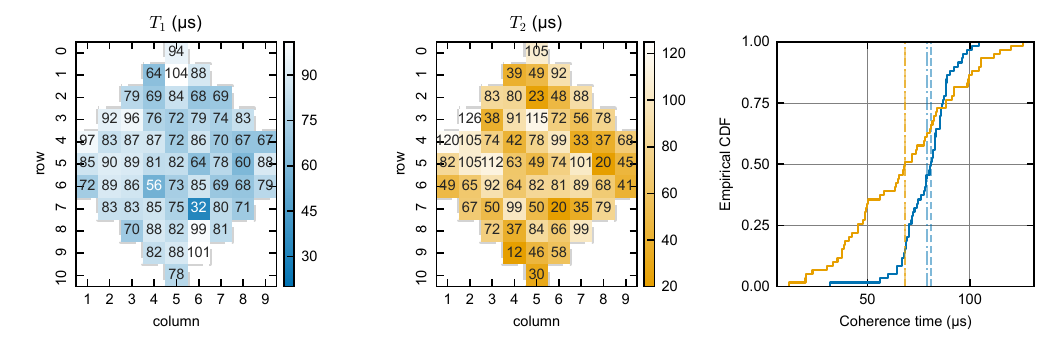}
    \caption{\textbf{iSWAP grid, coherence}. \textbf{a}: T1 and \textbf{b}: T2 of the iSWAP grid measure with interleaved population decay experiment and echo experiment. We measure the following statistical quantities for $T_1$: mean: $80$ $\mu$s, median: $81$ $\mu$s, std:  $12$ $\mu$s and for for $T_2$: mean: $68$ $\mu$s, median: $27$ $\mu$s, std:  $47$ $\mu$s      }
    \label{fig:supp_iSWAP_coherences}
\end{figure}

\begin{figure}[h]
    \centering
    \includegraphics{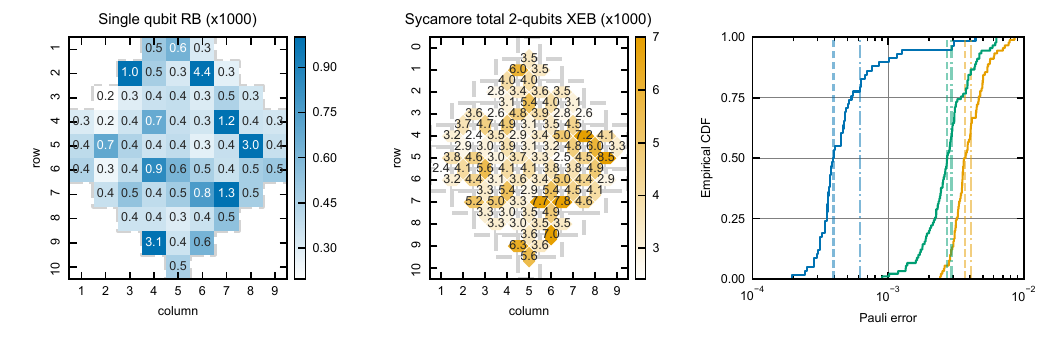}
    \caption{\textbf{iSWAP grid, Randomized Benchmarking}. \textbf{a}: Single qubit Pauli error measured through randomized benchmarking. \textbf{b}: Two qubit XEB Pauli error measured with XEB. Here we report the total error that include the single qubit error. \textbf{c}: Empirical cumulative histogram of the Pauli error rates across the chip. The green represents the \textit{inferred} error of the CZ where the single qubit gate error (blue) has been subtracted from the total XEB error (yellow). }
    \label{fig:supp_iSWAP_rb}
\end{figure}

\begin{figure}[h]
    \centering
    \includegraphics{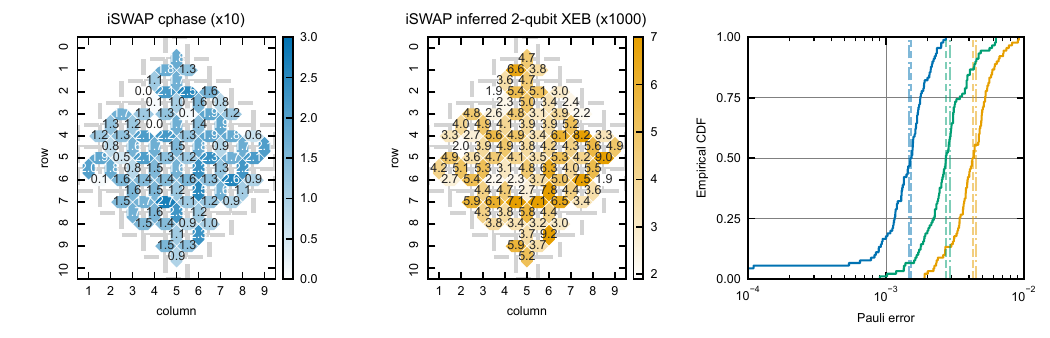}
    \caption{\textbf{iSWAP grid, iSWAP Benchmarking}. \textbf{a}: iSWAP cphase (in radians) measured using MEADD~\cite{gross_characterizing_2024}. \textbf{b}: Two qubit inferred XEB Pauli error for the iSWAP gate calculated by adding $\phi^2/16$ to the inferred Pauli error of the Sycamore gate where $\phi$ is the measured cphase.  \textbf{c}: Emperical cumulative histogram of the Pauli error rates across the chip. The blue is the measured cphase, 
    the green is the Sycamore inferred Pauli error, and the orange is the iSWAP inferred Pauli error. }
    \label{fig:supp_iSWAP_cphase}
\end{figure}

\begin{figure}[h]
    \centering
    \includegraphics{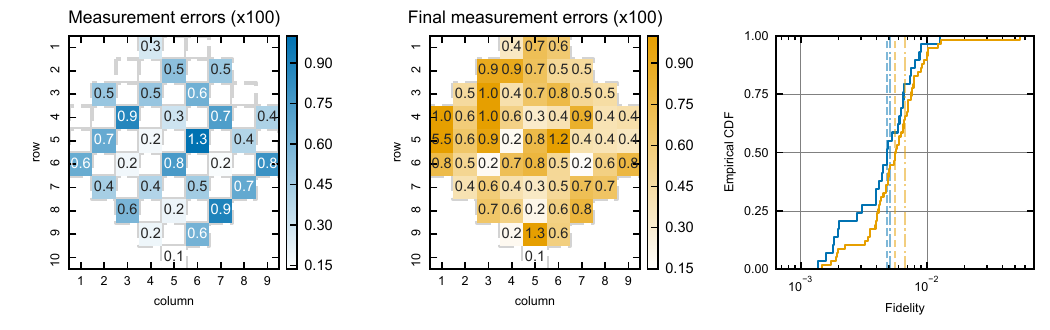}
    \caption{\textbf{iSWAP grid, Readout Benchmarking}. \textbf{a}: Average readout error during a mid-cycle measurement round using randomized state readout benchmarking. \textbf{b}: Same experiment for the final round of the memory experiment where all qubits are measured. \textbf{c}: Empirical cumulative histogram of the readout errors for both mid-circuit and termination measurements. }
    \label{fig:supp_iSWAP_readout}
\end{figure}

\begin{figure}[h]
    \centering
    \includegraphics{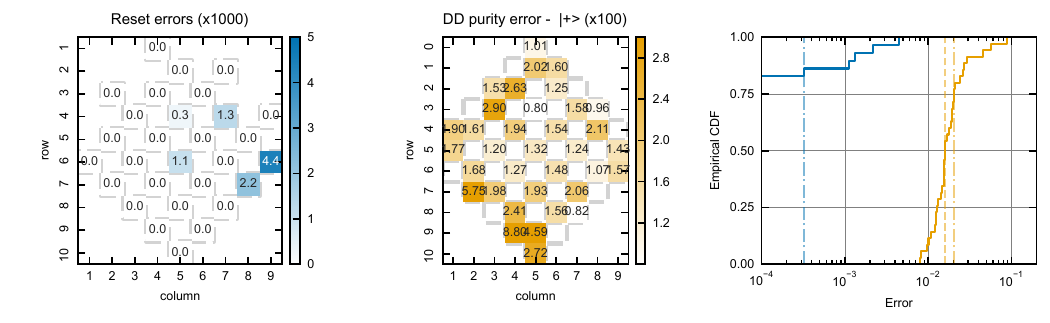}
    \caption{\textbf{iSWAP implementation, Reset and Dynamical Decoupling}: \textbf{A} Error (residual one state) after a reset on a measurement qubit. Note that this measurement used fewer statistics than the hexagonal and walking reset benchmarks, and as a result, many qubits report 0 reset errors. In our Pauli simulations, we have replaced these qubits with the median reset error value of all other qubits measured. We've found that this change makes little difference in the simulation results, which is expected given that the reset is a small contributor to the $1/\Lambda$ budget. \textbf{B} Dynamical decoupling purity error when initializing in $\ket{+}$ \textbf{C} Empirical cumulative histogram of the Reset and DD error rates. }
    \label{fig:supp_iSWAP_reset}
\end{figure}

In this subsection, we report all component benchmark measured for the iSWAP implementation. This implementation was performed using the 72-qubits Willow device also used in \cite{google_quantum_ai_quantum_2025}. ~\fig{supp_iSWAP_coherences} shows the coherence times of the grid used. ~\fig{supp_iSWAP_rb} shows the raw gate benchmarks with single qubit RB and two-qubits XEB (without the c-phase accounted for). ~\fig{supp_iSWAP_cphase} we report the measured c-phases. ~\fig{supp_iSWAP_readout} shows the readout benchmark for the mid-circuit and terminal measurements. ~\fig{supp_iSWAP_reset} shows the idling error during the measurement cycles (or error during the Dynamical Decoupling sequence). ~\fig{supp_iSWAP_reset} shows reset errors. All the component benchmarks are summarized in table~\ref{tab:iSWAP component_benchmarks}.

\begin{table}[h]
  \centering
  \begin{tabular}{|c|c|c|c|c|c|c|c|c|c|c|c|}
  \hline
   & count & mean & median & std & q1 & q3 & min & max & IQR & low outliers & high outliers \\
  \hline
  T1 (\SI{}{\micro\second}) & 59 & 79.32 & 80.86 & 11.98 & 71.02 & 86.78 & 31.58 & 104.38 & 15.76 & 1 & 0  \\
  T2 (\SI{}{\micro\second}) & 59 & 68.30 & 68.48 & 27.45 & 46.69 & 88.49 & 11.52 & 125.82 & 41.80 & 0 & 0  \\
  Single Qubit RB ($\times 10^3$) & 58 & 0.62 & 0.39 & 0.71 & 0.35 & 0.53 & 0.20 & 4.39 & 0.18 & 0 & 7  \\
  Sycamore two qubit XEB (total $\times 10^3$) & 90 & 4.07 & 3.67 & 1.24 & 3.21 & 4.65 & 2.42 & 8.51 & 1.43 & 0 & 5  \\
  Sycamore two qubit XEB (inferred $\times 10^3$) & 90 & 2.92 & 2.72 & 1.15 & 2.23 & 3.38 & 0.90 & 6.26 & 1.16 & 0 & 5  \\
   iSWAP cphase (Rad $\times 10^1$) & 90 & 1.46 & 1.52 & 0.57 & 1.16 & 1.77 & 0.00 & 2.71 & 0.61 & 5 & 1   \\
  iSWAP two qubit XEB (inferred $\times 10^3$) & 90 & 4.47 & 4.28 & 1.52 & 3.43 & 5.24 & 1.90 & 9.17 & 1.81 & 0 & 3  \\
  Mid-cycle readout ($\times 10^2$) & 29 & 0.51 & 0.49 & 0.26 & 0.31 & 0.65 & 0.14 & 1.28 & 0.35 & 0 & 1  \\
  Mid-cycle reset ($\times 10^3$) & 29 & 0.32 & 0.00 & 0.92 & 0.00 & 0.00 & 0.00 & 4.44 & 0.00 & 0 & 5  \\
  Final measurement ($\times 10^2$)& 58 & 0.67 & 0.56 & 0.68 & 0.41 & 0.75 & 0.15 & 5.47 & 0.34 & 0 & 2  \\
  Dynamical Decoding (T2) purity ($\times 10^2$) & 35 & 2.06 & 1.60 & 1.50 & 1.29 & 2.04 & 0.80 & 8.80 & 0.75 & 0 & 3  \\
  \hline
  \end{tabular}
  \label{tab:iSWAP component_benchmarks}
  \caption{\textbf{iSWAP implementation benchmarking.} Statistical description of the component benchmarks with Pauli error rates. All values are in units of Pauli error unless otherwise stated.}
\end{table}

\subsection{Detection analysis}
\begin{figure}[h]
    \centering
    \includegraphics{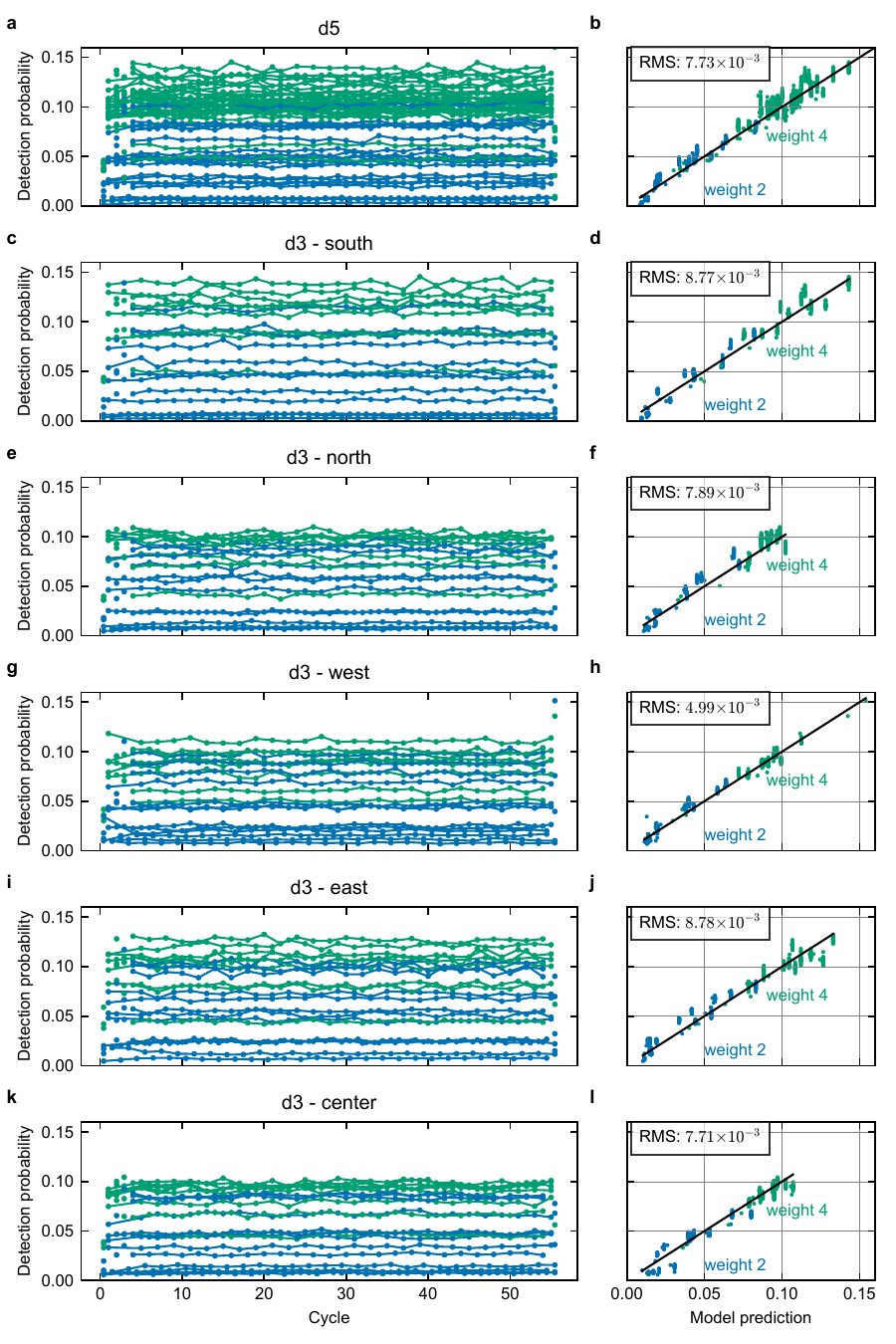}
    \caption{\textbf{iSWAP implementation, Detection Fraction analysis}. Left panels show the average detection fraction of all experiment in the Z basis. The right panels shows the correlation of the experimental detection probabilities with the models predictions.}
    \label{fig:supp_iswap_det_frac}
\end{figure}

In this section, we present a detailed analysis of the detection events experimentally collected using the iSWAP implementation. We mainly focus on the distance 5 and distance 3 iSWAP circuits with 54 cycles to confirm that our Pauli models properly predict the experimental data. Our first step is to look at the detection fraction (of first statistical moment) of the detection events. In the first column of ~\fig{supp_iswap_det_frac}, we show the average of each detectors over 20 different initial bitstrings and 10k repetitions per experiments. We see a clear separation between the boundary (weight 2) and the bulk (weight 4) detectors. 
In the second column of ~\fig{supp_iswap_det_frac}, we show the correlation of these detection fraction from simulations. We also present the RMS of the measured detection probabilities with the Pauli error model constructed from the benchmarks of the previous section \ref{subsec:supp_iSWAP_benchmarks}. We see that even though the benchmarks have been taken in the context of the distance-5 memory, the Pauli error model still capture the detection probabilities for the embedded distance-3 codes.

\begin{figure}[h]
    \centering
    \includegraphics{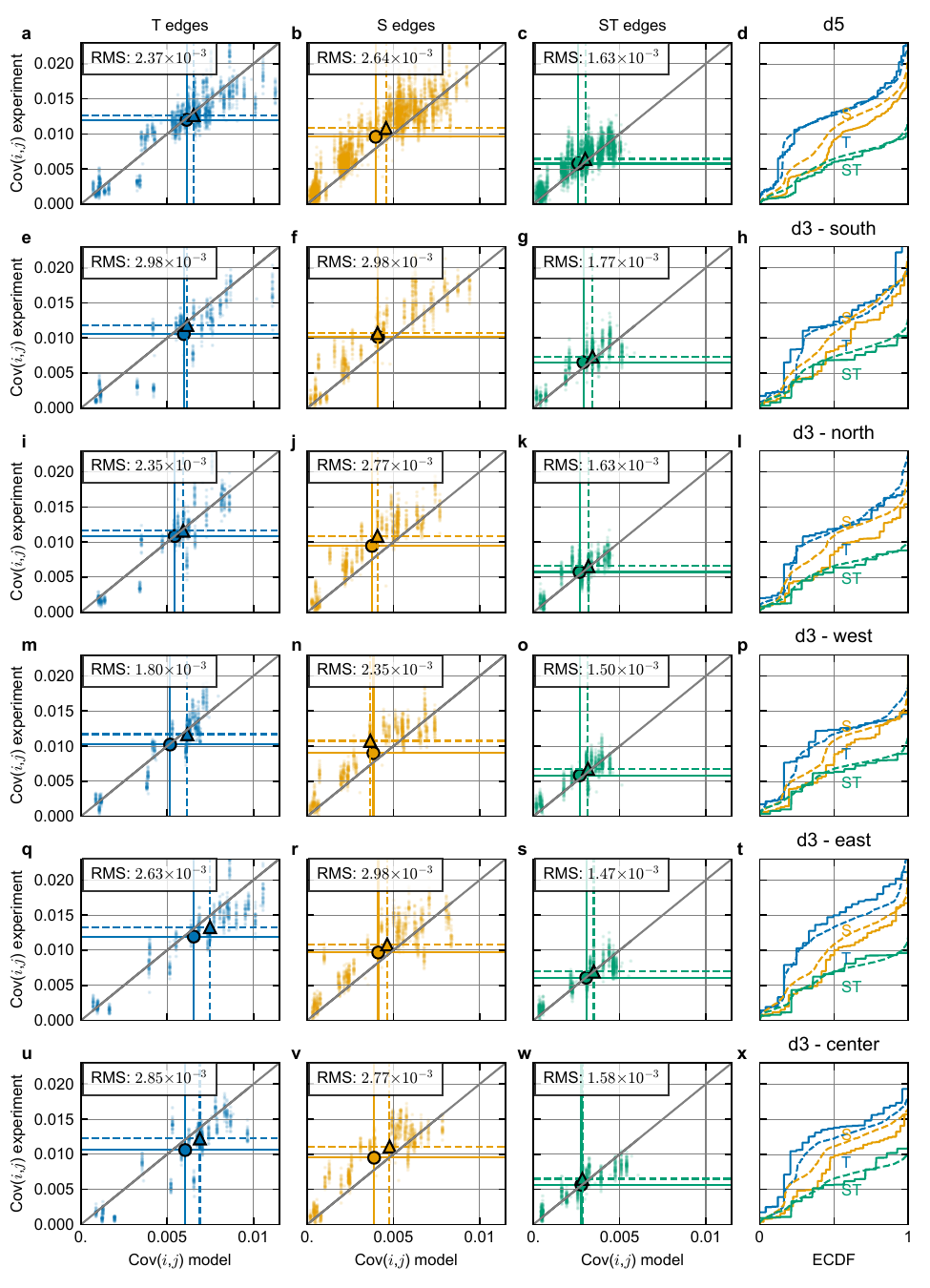}
    \caption{\textbf{iSWAP implementation, Detector covariance analysis -  regular edges}. \textbf{a column}. Space-like (S) edges. \textbf{b column}: Timelike (T) edges. \textbf{c column}: spacetimelike (ST) edges. \textbf{d column}: Cumulative histogram of the edges.}
    \label{fig:supp_iswap_covariance}
\end{figure}

    \section{Decoding results}
In this section, we list the logical error rates and error suppression factor $\Lambda_{35}$ for all codes demonstrated in this work. In the main text, we have systematically quoted the $\Lambda_{35}$ associated with the Harmony decoder and RL-optimized decoder prior, since this combination achieves the best logical performance.

We compare two decoders: sparse blossom \cite{higgott_sparse_2023} and Harmony \cite{shutty_efficient_2024} (an ensemble of 50 sparse blossom decoders). We also compare two methods for configuring the decoder prior: (i) the SI1000 error model \cite{Gidney_2021_honeycomb}, inspired by the typical hierarchy of error rates in superconducting qubits. This error model is agnostic of the device and the decoder. (ii) The reinforcement learning-based approach \cite{sivak_optimization_2024} that optimizes the decoder prior to reduce the logical error rate. In this approach, we adopt the hyperparameters from Ref.~\cite{sivak_optimization_2024} and train the prior on the ``calibration dataset'' of duration 16 cycles, separately for each code distance, patch location, and basis. The learned parameters of the prior are then extrapolated using the time-invariance assumption to decode all datasets up to 56 cycles. During training, we use the sparse blossom decoder, even when the final decoding is done with Harmony, as it was shown in \cite{sivak_optimization_2024} that the learned priors generalize across these two decoders.

\begin{table}[h!]
\begin{tabular}{|l|l|l|l|l|l|l|l|l|}
\hline 
decoder & prior & $\epsilon_{{3,X}}\times 10^3 $ & $\epsilon_{{3,Z}} \times 10^3$ & $\epsilon_{{3}}\times 10^3$ & $\epsilon_{{5,X}}\times 10^3$ & $\epsilon_{{5,Z}}\times 10^3$ & $\epsilon_{{5}}\times 10^3$ & $\Lambda_{{35}}$  \\ \hhline{|=|=|=|=|=|=|=|=|=|}
sparse blossom & SI1000 & $6.58(5)$ & $6.78(4)$ & $6.68(3)$ & $3.03(4)$ & $3.37(3)$ & $3.20(3)$ & $2.09(2)$ \\ \hline 
sparse blossom & RL & $6.08(4)$ & $6.29(3)$ & $6.19(3)$ & $2.89(4)$ & $3.22(3)$ & $3.06(2)$ & $2.02(2)$ \\ \hline 
harmony & SI1000 & $5.95(5)$ & $6.19(3)$ & $6.07(3)$ & $2.58(5)$ & $2.93(3)$ & $2.76(3)$ & $2.20(2)$ \\ \hline 
harmony & RL & $5.75(4)$ & $5.85(3)$ & $5.80(2)$ & $2.53(5)$ & $2.87(3)$ & $2.70(3)$ & $2.15(2)$ \\ \hline 
\end{tabular}
\caption{\textbf{Decoding results: hexagonal implementation}}
\end{table}

\begin{table}[h!]
\begin{tabular}{|l|l|l|l|l|l|l|l|l|}
\hline 
decoder & prior & $\epsilon_{{3,X}}\times 10^3 $ & $\epsilon_{{3,Z}} \times 10^3$ & $\epsilon_{{3}}\times 10^3$ & $\epsilon_{{5,X}}\times 10^3$ & $\epsilon_{{5,Z}}\times 10^3$ & $\epsilon_{{5}}\times 10^3$ & $\Lambda_{{35}}$  \\ \hhline{|=|=|=|=|=|=|=|=|=|}
sparse blossom & SI1000 & $14.9(5)$ & $13.0(3)$ & $13.9(3)$ & $8.2(4)$ & $7.9(2)$ & $8.0(2)$ & $1.74(6)$ \\ \hline 
sparse blossom & RL & $12.1(5)$ & $12.1(2)$ & $12.1(3)$ & $7.5(4)$ & $7.5(2)$ & $7.5(2)$ & $1.62(6)$ \\ \hline 
harmony & SI1000 & $14.2(5)$ & $12.4(2)$ & $13.3(3)$ & $7.6(4)$ & $7.2(2)$ & $7.4(2)$ & $1.80(6)$ \\ \hline 
harmony & RL & $11.8(5)$ & $11.8(2)$ & $11.8(3)$ & $7.0(4)$ & $7.1(2)$ & $7.0(2)$ & $1.69(6)$ \\ \hline 
\end{tabular}
\caption{\textbf{Decoding results:  walking implementation}}
\end{table}

\begin{table}[h!]
\begin{tabular}{|l|l|l|l|l|l|l|l|l|}
\hline 
decoder & prior & $\epsilon_{{3,X}}\times 10^3 $ & $\epsilon_{{3,Z}} \times 10^3$ & $\epsilon_{{3}}\times 10^3$ & $\epsilon_{{5,X}}\times 10^3$ & $\epsilon_{{5,Z}}\times 10^3$ & $\epsilon_{{5}}\times 10^3$ & $\Lambda_{{35}}$  \\ \hhline{|=|=|=|=|=|=|=|=|=|}
sparse blossom & SI1000 & $12.3(1)$ & $10.32(6)$ & $11.33(6)$ & $8.80(9)$ & $6.6(1)$ & $7.71(7)$ & $1.47(2)$ \\ \hline 
sparse blossom & RL & $11.0(1)$ & $9.78(4)$ & $10.41(5)$ & $7.8(1)$ & $6.07(6)$ & $6.94(8)$ & $1.50(2)$ \\ \hline 
harmony & SI1000 & $11.6(1)$ & $9.90(5)$ & $10.75(6)$ & $8.0(1)$ & $6.1(1)$ & $7.07(8)$ & $1.52(2)$ \\ \hline 
harmony & RL & $10.8(1)$ & $9.45(5)$ & $10.15(6)$ & $7.3(1)$ & $5.74(8)$ & $6.50(7)$ & $1.56(2)$ \\ \hline 
\end{tabular}
\caption{\textbf{Decoding results: iSWAP implementation}}
\end{table}
    \section{Simulation of the gate frequency optimization \label{SM:hex_vs_standard_sims}}

\begin{figure*}[h]
    \centering
    \includegraphics{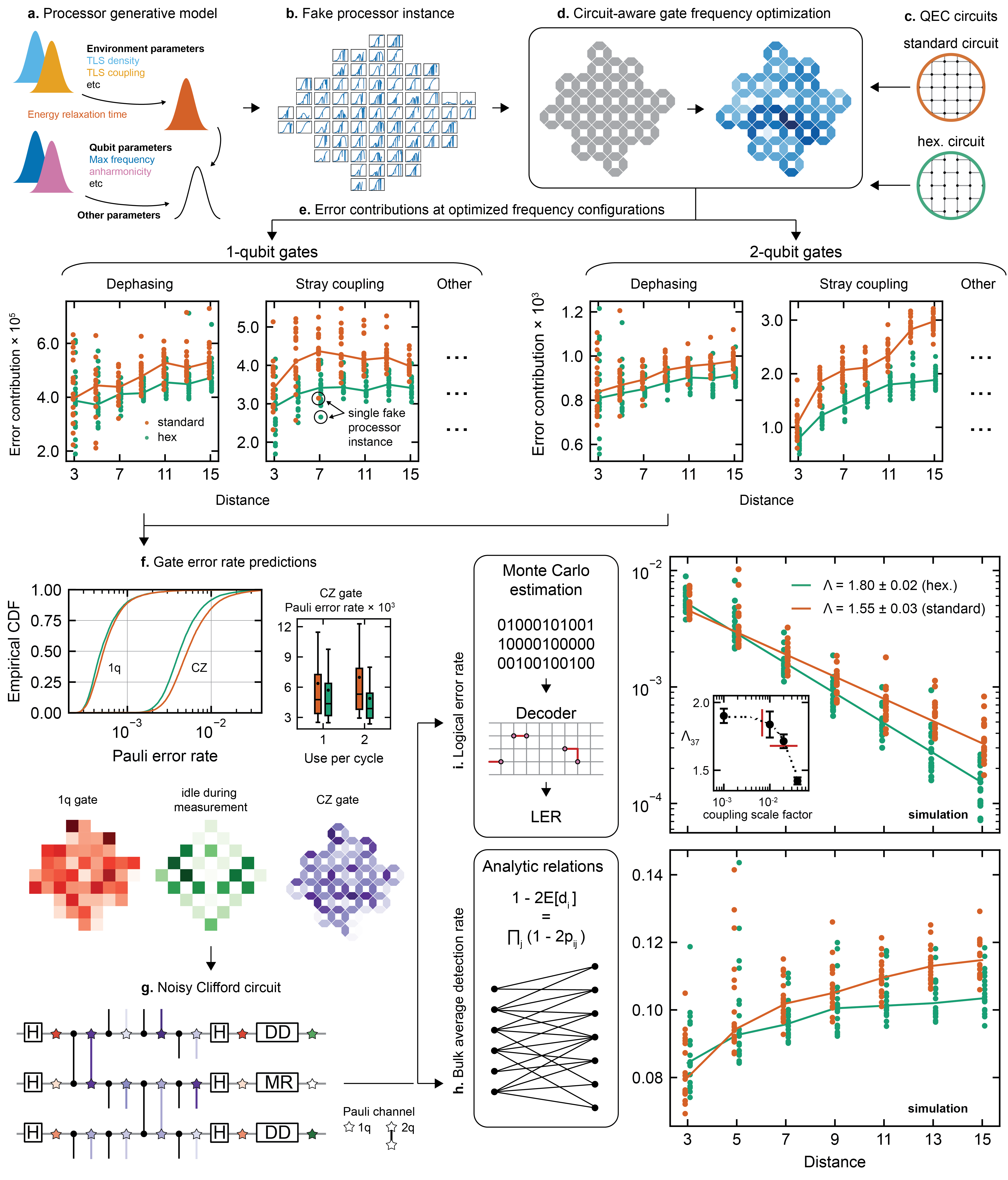}
    \caption{\textbf{Simulation of the gate frequency optimization for the standard vs hexagonal surface code circuits}. See Section~\ref{SM:hex_vs_standard_sims} for details.}
    \label{fig:supp_hex_vs_standard_sims}
\end{figure*}

In this section, we describe the simulation of the gate frequency optimization for the standard vs hexagonal surface code circuits. The goal of this simulation is to demonstrate that removing one coupler from the qubit grid unit cell relaxes the constraints on the frequency selection algorithm, which leads to improved performance.

The simulations begin by producing fake processors in software, whose properties are statistically indistinguishable from the real processors. To this end, we use a generative model parametrized as a Bayesian network \cite{klimov_snake_2020}.
As illustrated in ~\fig{supp_hex_vs_standard_sims}\textbf{a}, in this model the root variables associated with different qubits and couplers of the processor are assumed to be statistically independent. These variables are related to architectural parameters of the device (e.g. the maximum frequency of its tuning range, the coupling efficiency to the neighbouring devices, the stray coupling strength, etc) and parameters of the local environment of the device (e.g. spectral density of the two-level system (TLS) defects, coupling strengths to these TLSs, etc). Other variables are then derived by propagating through the conditional dependencies defined by the Bayesian network. For example, the fake characterization data for the energy relaxation time as a function of qubit frequency, produced in this manner, is shown in Fig.~\ref{fig:supp_hex_vs_standard_sims}(b). Each sample fetched from the generative model has slightly different properties, which mimics the variation in real devices. In this study, we sample $20$ such fake processor instances for each code distance.

The sampled processors initially have square connectivity with 4 neighbours per qubit. To ensure a fair comparison to hexagonal connectivity with 3 neighbours per qubit, instead of sampling a new set of processors we modify the existing processors by removing one coupler per qubit and scaling the qubit-qubit coupling efficiency across the removed coupler by a factor $10^{-3}$. This procedure leads to a new set of processors with effectively hexagonal connectivity, where each processor is otherwise identical to its twin with the square connectivity. 

For the superconducting processors, it was previously established that most of the error mechanisms that corrupt the gate fidelity have nontrivial frequency dependence. For example, CZ gates between pairs of neighbouring qubits are enacted by bringing them into a certain resonance condition. If the spectral location of this resonance happens to coincide with a location of the TLS, then the gate fidelity can strongly degrade. Other frequency-dependent error mechanisms include stray interaction leading to gate crosstalk, flux sensitivity leading to dephasing, etc. Therefore, it is crucial to optimize the frequency configuration for all gates in a quantum circuit to achieve the highest performance, see ~\fig{supp_hex_vs_standard_sims}\textbf{d}. Our model-based framework for such optimization is described in detail in Ref.~\cite{klimov_optimizing_2024}. Here, we use the Snake optimizer \cite{klimov_snake_2020} with local optimization of scope 3 and no additional healing. We launch Snake from 20 different initial seeds for each fake processor and select the solution that achieves the minimal expected QEC detection rate for a given circuit, either hexagonal or standard, see ~\fig{supp_hex_vs_standard_sims}\textbf{c}. This circuit-aware frequency optimization is done for the standard circuit on the square-connectivity grid and for the hexagonal circuit on the twin hexagonal-connectivity grid.  

When the frequency optimization is completed, we inspect various gate error contributions achieved in the final frequency configuration. As shown in ~\fig{supp_hex_vs_standard_sims}\textbf{e}, hexagonal connectivity allows to suppress the effect of several different error contributions. For example, since each qubit effectively has only 3 neighbors, the potential for spurious frequency collisions is reduced. This allows to bunch the qubit frequencies in the narrower bandwidth close to the maximum of the tuning range to suppress the sensitivity to dephasing from flux noise. We observe a clear trend in reduced contribution from dephasing in the hexagonal circuit due to this effect. Additionaly, the contribution from stray interaction with neighbouring qubits is reduced by a factor close to the $4/3$ reduction of the number of neighbours. We find that some other error contributions (not shown) remain similar in the two cases. However, the overall effect of reduced connectivity allows to achieve lower gate error rates.

The empirical cumulative distribution function (CDF) of the predicted gate Pauli errors for single- and two-qubit gates over all sampled distance-15 fake processors is shown in ~\fig{supp_hex_vs_standard_sims}\textbf{f}. For our choice of the processor generative model, we find that the mean single-qubit gate error rate is suppressed in the hexagonal connectivity by $3.4\%$, and two-qubit gate error rate is suppressed by $21.0\%$. Moreover, since our frequency optimization framework is circuit-aware, it was able to prioritize the CZ gates that are used twice per cycle in the hexagonal circuit, improving those on average by $30.0\%$ compared to the standard circuit, while those that are used once per cycle are improved on average by only $10.5\%$. In the right panel of ~\fig{supp_hex_vs_standard_sims}\textbf{f} illustrating this effect, the box extends from the first to the third quartile of the data, with a line at the median and a circle at the mean. The whiskers indicate the lower and upper $10\%$ of the data.

Once the predictions of the gate error rates are available, we proceed to simulate the QEC circuit under the Pauli noise model. We model the single- and two-qubit gate errors with the single- and two-qubit depolarizing channels respectively, which are appended after the ideal gate. In addition, we include depolarizing channels on the data qubits during their idle time when the measure qubits are being measured and reset. The frequency optimization procedure in this study did not take into account the error channels affecting the qubit readout. We heuristically included a uniform readout error probability of $8\times10^{-3}$ on all qubits. In the Pauli simulation, this probability was used to stochastically flip the measurement outcomes. The resulting noisy Clifford circuit can be used for various downstream tasks. In particular, we use it to predict the bulk average detection rate (BADR), shown in ~\fig{supp_hex_vs_standard_sims}\textbf{h}, and the logical error rate (LER), shown in ~\fig{supp_hex_vs_standard_sims}\textbf{i}. 

We compute BADR by first employing Stim \cite{gidney_stim_2021} to automatically produce the detector error model (DEM) for the noisy circuit, and then using the analytic relation between the detection rate of detector $i$ and error probabilities $p_{ij}$ for independent error mechanisms $j$ in the DEM that affect this detector $i$:
\begin{align}
    1-2\,{\mathbb E}[d_i] = \prod_{j: \langle ij\rangle\in\rm DEM} (1-2p_{ij})
\end{align}
We further average the detection rate over all bulk detectors to obtain the BADR. We find that for both circuits BADR increases with code distance, indicating that the frequency allocation problem becomes more difficult for larger processors. This trend is also observable in the gate error contributions in ~\fig{supp_hex_vs_standard_sims}\textbf{e}, especially the contribution from the stray coupling. We emphasize that this increase of the average detection rate with distance cannot be explained by the dilution of the role of the boundary in larger code patches, because we exclude the boundary detectors from this averaging. Eventually, the detection rate appears to saturate, and the saturation level is lower for the hexagonal circuit, consistent with it having lower gate error rates. We also note that a similar saturation effect was previously observed for the parallel XEB fidelity as a function of the processor size in the scaling simulations done in Ref.~\cite{klimov_optimizing_2024}. It remains an open question to what extent this effect can be mitigated through improved frequency optimization algorithms.

Although BADR and its breakdown into various error contributions gives us valuable information about the system, the ultimate measure of the processor quality from the QEC perspective is the LER. To estimate LER in this simulation, we sample $5\times10^6$ shots from the noisy circuit and decode them using correlated matching, a fast decoder compatible with the real-time operation \cite{higgott_sparse_2023, google_quantum_ai_quantum_2025}. We find that LER of the fake processors sampled from the same generative model can vary by almost an order of magnitude. It is possible that this variance can be tightened by performing multiple rounds of healing to locally fix the outlier qubits that were not properly optimized. By fitting the mean LER over all fake processors versus code distance to the typical scaling law ${\rm LER}(d)\propto \Lambda^{-(d+1)/2}$, we obtain the error suppression factor $\Lambda$ for the standard and hexagonal circuits of $1.55\pm 0.03$ and $1.80\pm 0.02$ respectively, equivalent to $16\%$ improved scaling for the hexagonal connectivity.

In this simulation, the qubit-qubit coupling across the removed coupler in the hexagonal grid was suppressed by a factor $10^{-3}$ compared to its value in the twin square grid. To estimate the fabrication tolerance on this parameter required to achieve the demonstrated improvement in $\Lambda$, we sweep the scaling factor for the qubit-qubit coupling, see ~\fig{supp_hex_vs_standard_sims}\textbf{i} inset. The scaling factor at which the coupling strength becomes comparable to the stray diagonal coupling is marked with a vertical red notch at $\sim 7\times 10^{-3}$, and the $\Lambda$ of the standard circuit is marked with a horizontal notch. Note that in this sweep of the coupling efficiency we only used fake processors of distance up to $7$, which results in slightly higher $\Lambda$ estimates than in the large-distance limit. For example, the standard circuit has $\Lambda = 1.68$ (horizontal red notch) when fitted only on distances up to $7$, and $1.55$ when fitted on distances up to $15$. It is likely due to the slow saturation of the physical error rates with code distance, mentioned earlier. Regardless of this, from the sweep results we conclude that reducing the qubit-qubit coupling across the removed coupler to the level of the stray diagonal coupling is sufficient to achieve the promised improvement of $\Lambda$.
\clearpage

    \clearpage
    \newpage
\onecolumngrid

\vspace{1em}
\begin{flushleft}
{\small \textbf{Google Quantum AI and Collaborators}}

\bigskip
{\small
\renewcommand{\author}[2]{#1$^\textrm{\scriptsize #2}$}
\renewcommand{\affiliation}[2]{$^\textrm{\scriptsize #1}$ #2 \\}

\newcommand{\corrauthora}[2]{#1$^{\textrm{\scriptsize #2}, \ddagger}$}
\newcommand{\corrauthorb}[2]{#1$^{\textrm{\scriptsize #2}, \mathsection}$}

\newcommand{\xGoogle}{\affiliation{1}{Google Quantum AI, Santa Barbara, CA 93117, USA}}

\newcommand{\xUCONN}{\affiliation{2}{Department of Physics, University of Connecticut, Storrs, CT, USA}}

\newcommand{\xUMass}{\affiliation{3}{Department of Electrical and Computer Engineering, University of Massachusetts, Amherst, MA}}

\newcommand{\xAU}{\affiliation{4}{Department of Electrical and Computer Engineering, Auburn University, Auburn, AL, USA}}

\newcommand{\Google}{1}
\newcommand{\UCONN}{2}
\newcommand{\UMass}{3}
\newcommand{\AU}{4}

\corrauthora{Alec Eickbusch}{\Google},
\corrauthora{Matt McEwen}{\Google},
\author{Volodymyr Sivak}{\Google},
\author{Alexandre Bourassa}{\Google},
\author{Juan Atalaya}{\Google},
\author{Jahan Claes}{\Google},
\author{Dvir Kafri}{\Google},
\author{Craig Gidney}{\Google},
\author{Christopher W.~Warren}{\Google},
\author{Jonathan Gross}{\Google},
\author{Alex Opremcak}{\Google},
\author{Nicholas Zobrist}{\Google},
% ...
\corrauthora{Alexis Morvan}{\Google}.

\bigskip

\xGoogle
\xUCONN
\xUMass
\xAU

{${}^\ddagger$ These authors contributed equally to this work.}

}
\end{flushleft}
    \bibliography{bib.bib}